\documentclass[twocolumn, trackchanges]{aastex7}
\usepackage{amsmath}
\usepackage{CJKutf8}
\usepackage{siunitx}

\usepackage{multirow}

\newcommand{\iso}[2]{\ensuremath{^{#2}\mathrm{#1}}}

\newcommand{\rev}[1]{#1}

\begin{document}

\title{ The Impact of the New \iso{Fe}{59} Decay Rates on \iso{Fe}{60} and
     \iso{Al}{26} Nucleosynthesis in Massive Stars. }

\author[0009-0003-9782-2501]{Bingyang Tan}
\affiliation{CAS Key Laboratory of Optical Astronomy, National Astronomical Observatories, Chinese Academy of Sciences, Beijing 100101, People's Republic of China}
\affiliation{School of Astronomy and Space Science, University of Chinese Academy of Sciences, Beijing 100049, People's Republic of China}
\email{tanby@nao.cas.cn}

\author[0000-0003-3646-9356]{Wenyu Xin}
\affiliation{Institute of Astronomy and Physics, Inner Mongolia University, Hohhot 010021, People's Republic of China}
\email[show]{xinwenyu16@mails.ucas.ac.cn}

\author[0009-0001-0604-072X]{Ruizheng Jiang}
\affiliation{CAS Key Laboratory of Optical Astronomy, National Astronomical Observatories, Chinese Academy of Sciences, Beijing 100101, People's Republic of China}
\affiliation{School of Astronomy and Space Science, University of Chinese Academy of Sciences, Beijing 100049, People's Republic of China}
\email{jiangrz@bao.ac.cn}

\author[0000-0002-8980-945X]{Gang Zhao}
\affiliation{CAS Key Laboratory of Optical Astronomy, National Astronomical Observatories, Chinese Academy of Sciences, Beijing 100101, People's Republic of China}
\affiliation{School of Astronomy and Space Science, University of Chinese Academy of Sciences, Beijing 100049, People's Republic of China}
\email[show]{gzhao@nao.cas.cn}

\author[0000-0002-6705-6303]{Koh Takahashi}
\affiliation{National Astronomical Observatory of Japan, National Institutes for Natural Science, 2-21-1 Osawa, Mitaka, Tokyo 181-8588, Japan}
\email{koh.takahashi@nao.ac.jp}


\correspondingauthor{Wenyu Xin, Gang Zhao}

\begin{abstract}
The diffuse $\gamma$-ray emission from short-lived radioactive \iso{Al}{26} and
\iso{Fe}{60} provides a direct probe of ongoing nucleosynthesis in the Galaxy.
However, theoretical models have long struggled to reproduce the observed
\iso{Fe}{60}/\iso{Al}{26} flux ratio, typically predicting values significantly
higher than constraints derived from INTEGRAL/SPI observations. In this work, we
investigate the impact of the recently measured, temperature-dependent stellar
$\beta^-$ decay rate of \iso{Fe}{59} on the nucleosynthesis of these isotopes.
We compute a grid of non-rotating massive star models ($14\text{--}80\,M_\odot$)
at solar metallicity using the \texttt{MESA} code, coupled with a rigorous
numerical resolution analysis. We find that the updated rate significantly
suppresses the net production of \iso{Fe}{60} by approximately $0.28$\,dex
($\sim 47\%$) compared to models using \texttt{LMP} theoretical rates, while
leaving \iso{Al}{26} yields virtually unchanged. This reduction is primarily
driven by the enhanced $\beta^-$ decay during convective carbon shell burning.
Integrating these yields over a standard Salpeter Initial Mass Function, we
predict a Galactic flux ratio of $\sim 0.18$, which is in excellent agreement
with the observed value of $0.184 \pm 0.042$. Furthermore, this ratio exhibits a
weak dependence on the IMF slope. Our results indicate that the updated nuclear
physics input significantly alleviates the long-standing \iso{Fe}{60}
overproduction problem, bringing theoretical predictions into much closer
alignment with current Galactic observations.

\end{abstract}



\section{Introduction}
\label{sec:intro}

Observations of diffuse $\gamma$-ray lines emitted from the decay of radioactive
isotopes \iso{Al}{26} (at 1.809 MeV) and \iso{Fe}{60} (at 1.173 and
1.332 MeV) provide a direct probe of ongoing nucleosynthesis in the Milky Way
\citep{1982ApJ...262..742M, 2004ESASP.552...45S, 2007A&A...469.1005W,
2020ApJ...889..169W, 2021PASA...38...62D}. With respective half-lives of 0.717
Myr for \iso{Al}{26} \citep{PhysRevC.61.015801, PhysRevC.97.065807} and 2.62 Myr
for \iso{Fe}{60} \citep{2009PhRvL.103g2502R, 2015PhRvL.114d1101W,
PhysRevC.95.055809}, these isotopes trace nucleosynthetic events occurring
within the last few million years.

All-sky map of the 1.809 MeV $\gamma$-ray emission suggests that massive stars
and their subsequent core-collapse supernovae (CCSNe) are the primary production
sites for \iso{Al}{26} \citep{1995A&A...298..445D,2004A&A...420.1033P}. As
summarized by \citet{2021PASA...38...62D}, while both isotopes \iso{Al}{26} and
\iso{Fe}{60} are synthesized in Asymptotic Giant Branch (AGB) and massive stars,
the contribution from AGB stars is generally considered minor and remains
subject to significant uncertainties inherent in current stellar models,
especially for \iso{Fe}{60}. In the context of massive stars, given that both
isotopes share similar astrophysical origins and comparable half-lives, the
$\iso{Fe}{60}/\iso{Al}{26}$ $\gamma$-ray flux ratio is expected to be largely
insensitive to the details of their Galactic distribution. Under the assumption
of a steady-state equilibrium, where the continuous production of these isotopes
is balanced by their radioactive decay, the observed ratio could be directly
compared to the nucleosynthesis yields from massive star models by converting
the mass yields with a factor of 26/60. \citet{2017ApJ...839L...9A} considered
the $\iso{Fe}{60}/\iso{Al}{26}$ yield ratio as a `robust observable' for massive
star models. However, systematic discrepancies remain between theoretical
predictions and the flux ratios observed by Spectrometer on the International
Gamma Ray Astrophysics Laboratory \citep[INTEGRAL/SPI, ][]{2003A&A...411L..63V},
and Reuven Ramaty High-Energy Solar Spectroscopic Imager \citep[RHESSI,
][]{2002SoPh..210....3L}.

The first observational constraint on Galactic \iso{Fe}{60} emission was derived
from RHESSI data \citep{2004ESASP.552...45S}, yielding an
$\iso{Fe}{60}/\iso{Al}{26}$ flux ratio of approximately 0.4. Subsequent
observations by INTEGRAL/SPI have refined this value, with reported flux ratios
ranging from 0.08 to 0.22, adapting different analysis methods
\citep{2007A&A...469.1005W, 2011ApJ...739...29B, 2015ApJ...801..142B}. The most
recent results from 15 years of SPI data is presented by
\citet{2020ApJ...889..169W}. The \iso{Fe}{60}/\iso{Al}{26} flux ratio of $(18.4
\pm 4.2)\%$ is determined using a specific parameterized spatial morphology
model (e.g., an exponential disk). However, because the spatial distribution of
\iso{Fe}{60} is not yet as well-constrained as that of \iso{Al}{26}, a broader
range of $0.2 \text{--} 0.4$ is estimated by taking into account the sensitivity
of the results to different spatial tracer templates. This broader range
reflects the systematic uncertainties inherent in deconvolving diffuse emission
from the Galactic plane. This observational range is consistent with the 
lower limit of 0.1 inferred from deep-sea sediments data \citep{2018PhRvL.121v1103F}. 
\rev{Notably, early predictions by \citet{1995ApJ...449..204T} ($\sim$0.16)
align well with these subsequent observations. However, a persistent discrepancy
has emerged in later theoretical studies, with contemporary models
systematically overproducing \iso{Fe}{60} relative to \iso{Al}{26}. }
Specifically, several widely used models predict flux ratios that
exceed observed values by a factor of 3 to 10 \citep{ 2007PhR...442..269W,
2016ApJ...821...38S, 2017ApJ...839L...9A}. In contrast, certain studies
\citep{2006ApJ...647..483L, 2013ApJ...764...21C} yield more consistent results
in the range of 0.1 to 0.3, which align more closely with the observational
constraints. Such a tension suggests that our understanding of the
nucleosynthesis of these two isotopes in massive stars is unsettled, and the
general overproduction of \iso{Fe}{60} remains a fundamental challenge in
theoretical predictions.

The predicted yields of \iso{Al}{26} and \iso{Fe}{60} are known to be sensitive
to several key stellar modeling uncertainties. In particular, the inclusion of
stellar rotation can fundamentally alter the nucleosynthetic output by inducing
internal mixing and enhancing mass-loss rates. For \iso{Al}{26}, the
contribution from stellar winds is critically dependent on these processes, as
rotationally induced mixing can dredge up synthesized material to the stellar
surface, where it is subsequently ejected before the supernova stage. Previous
studies have extensively explored how different treatments of rotation and
mass-loss prescriptions impact the resulting \iso{Fe}{60} and \iso{Al}{26}
yields \citep{2012A&A...537A.146E, 2018ApJS..237...13L, 2021ApJ...923...47B,
2025ApJ...991...21F}. While adjustments to these modeling parameters can
sometimes improve the agreement with Galactic observations, they often introduce
further complexities and degeneracies within the stellar evolutionary tracks. 

However, a more fundamental and perhaps critical source of this discrepancy may
lie in the nuclear physics inputs, as \citet{2007PhR...442..269W} suggested, the
unmeasured cross sections governing the production of \iso{Fe}{60} could be a
significant source of uncertainty, for instance, the $\beta^-$ decay rate of
\iso{Fe}{59}. In the C and Ne burning shells of massive stars where \iso{Fe}{60}
is primarily synthesized, \iso{Fe}{59} acts as a pivotal branching point in
s-process pathways. The competition between neutron capture on \iso{Fe}{59} to
form \iso{Fe}{60} and its $\beta^-$ decay back to \iso{Co}{59} directly
influences the final yields of \iso{Fe}{60}. For decades, stellar models have
relied on theoretical rates from \citet{2001ADNDT..79....1L} (hereafter LMP),
which were based on large-scale shell-model calculations. 

In recent years, a breakthrough experiment by \citet{2021PhRvL.126o2701G}
provided the first experimental constraints on the key $\beta^-$ decay
transitions of \iso{Fe}{59}. Their results indicate that at temperatures
relevant to C and Ne shell burning ($T \gtrsim 1$ GK), the stellar decay rate is
nearly three times higher than the LMP prediction, \rev{directly suppressing
\iso{Fe}{60} production. Indeed,} \citet{2021PhRvL.126o2701G} also performed a
stellar model test for an 18~$M_{\odot}$ star, finding that the new rate reduces
the \iso{Fe}{60} yield by 40\% compared to the LMP rate. To fully evaluate the
impact on the Galactic \iso{Fe}{60}/\iso{Al}{26} flux ratio, however, a more
extensive grid of massive star models across a wide mass range is essential
\rev{to isolate the impact of this decay uncertainty on the final yields.}

\rev{Regarding the aforementioned branching-point competition, a recent
measurement by \citet{2024NatCo..15.9608S} reported an enhancement by a factor
of 1.6 to 2.1 in the \iso{Fe}{59} neutron capture rate, which would conversely
promote \iso{Fe}{60} production. While such updates should ideally be integrated
into stellar models concurrently, implementing multiple significant adjustments
simultaneously makes it difficult to disentangle the individual nucleosynthetic
impact of each physical change. Therefore, in this work, to facilitate a clear
comparison with previous theoretical studies, we retain the standard LMP rate
for the neutron capture channel. This strategy allows us to focus exclusively on
the consequences of the enhanced $\beta^-$ decay rate, which itself represents a
substantial threefold revision and a key mechanism for alleviating the
\iso{Fe}{60} overproduction problem.}

In this paper, we systematically investigate the impact of this updated
\iso{Fe}{59} decay rate on the nucleosynthesis of \iso{Fe}{60} in massive stars
and the resulting Galactic \iso{Fe}{60}/\iso{Al}{26} flux ratio. Utilizing the
Modules for Experiments in Stellar Astrophysics \citep[MESA, version
12115;][]{2011ApJS..192....3P, 2013ApJS..208....4P, 2015ApJS..220...15P,
2018ApJS..234...34P, 2023ApJS..265...15J}, we computed a comprehensive grid of
stellar models with initial masses from 14 to 80 $M_{\odot}$ at solar
metallicity. We compare two distinct sets of models: the "DR" series using
default LMP rates and the "NR" series incorporating the new experimental rate
from \citet{2021PhRvL.126o2701G}. By integrating these yields over a standard
Initial Mass Function (IMF), we quantify how the new nuclear data reconciles
theoretical predictions with the latest INTEGRAL observations.

This paper is organized as follows. Section~\ref{sec:methods} details our
numerical framework, including the pre-supernova stellar models, the treatment
of core-collapse explosions, and the integration procedure used to calculate the
Galactic \iso{Fe}{60}/\iso{Al}{26} flux ratio. In
Section~\ref{sec:results}, we present the basic evolutionary properties of our
models and analyze the primary nucleosynthesis yields, specifically highlighting
the physical mechanism of \iso{Fe}{60} suppression via a representative
30~$M_{\odot}$ case study. Section~\ref{sec:discussion} examines the broader
implications of these results, addressing the mass-dependent role of fallback,
comparisons with current INTEGRAL observations, and the impact of systematic
uncertainties. Finally, our main conclusions are summarized in
Section~\ref{sec:conclusion}.

\section{Numerical Methods}\label{sec:methods}

To quantify the impact of updated nuclear data on the Galactic
\iso{Fe}{60}/\iso{Al}{26} $\gamma$-ray flux ratio, we perform a multi-stage
numerical investigation. Our methodology bridges microscopic nuclear reaction
rates with macroscopic stellar populations through three steps: (1) the
construction of robust pre-supernova (pre-SN) models incorporating a refined
nuclear network; (2) the simulation of the core-collapse supernova (CCSN)
explosion via a parameterized thermal bomb to account for explosive
nucleosynthesis; and (3) the integration of these individual yields over the
initial mass function (IMF). In the following subsections, we detail the input
physics, the implementation of the new \iso{Fe}{59} decay rate, and the
resolution convergence tests that ensure the robustness of our nucleosynthetic
yields. Finally, we describe our CCSN models and procedure for calculating the
Galactic \iso{Fe}{60}/\iso{Al}{26} $\gamma$-ray flux ratio from our model grid.

\subsection{Pre-supernova Models}
We employ the \texttt{MESA} stellar evolution code (version r12115) to simulate
the evolution and nucleosynthesis of massive stars, from the zero-age main
sequence (ZAMS) through the onset of core-collapse supernova (CCSN) phase. The
pre-supernova models are evolved until any location within the iron core reaches
an infall velocity of $1000\,\text{km\,s}^{-1}$. Our setup is based on the
\texttt{25M\_pre\_ms\_to\_core\_collapse} test suite provided in the
\texttt{MESA} distribution.

\subsubsection{Basic Input Physics}
We simulate a grid of massive star models with initial masses ranging from $14$
to $80\,M_\odot$. This range covers the primary contributors to the Galactic
nucleosynthesis of \iso{Al}{26} and \iso{Fe}{60}
\citep{2006ApJ...647..483L,2007PhR...442..269W,2021PASA...38...62D}. Stars below
$14\,M_\odot$ are excluded as they yield significantly lower amounts of these
isotopes. While stars exceeding $80\,M_\odot$ could also contribute to the
Galactic \iso{Al}{26} budget through intense Wolf-Rayet (WR) wind activity
\citep{2005A&A...429..613P}, we truncate our grid at this limit because such
massive objects are prone to direct black hole formation, which potentially
retains their nucleosynthetic products. The potential impact of this limited
mass range on the integrated Galactic flux ratio is addressed in Section
\ref{subsubsec: stellar sources}.

All models adopt an initial metallicity of $Z = 0.02$ with the solar abundance
distribution from \citet{1998SSRv...85..161G}. The initial helium mass fraction
is set to $Y=2Z+0.24$ and the hydrogen mass fraction is $X=1-Y-Z$.
For mass loss, the models are evolved with the \texttt{Dutch} prescriptions
which includes \citet{1988A&AS...72..259D} for cool stars,
\citet{2001A&A...369..574V} for hot hydrogen-rich stars, and
\citet{2000A&A...360..227N} for Wolf-Rayet stars. We used a reduction factor of
\texttt{Dutch\_scaling\_factor} = 0.8, consistent with
\citet{2016ApJS..227...22F} and \citet{2001A&A...373..555M}.

Convective mixing is treated using the Ledoux criterion with a mixing length
parameter $\alpha_{\rm mlt} = 1.5$. We employ the diffusive exponential
overshoot scheme \citep{2000A&A...360..952H} with MESA parameters $f = 0.01$ and
$f_0 = 0.004$. These values, along with a semi-convection efficiency of
$\alpha_{\rm semi} = 0.01$, are chosen to ensure our stellar structures are
consistent with established grids in the literature
 \citep{2016ApJS..227...22F, 2023ChPhC..47c4107X, 2025arXiv250211012X},
thereby providing a robust baseline for evaluating the impact of nuclear decay
rate uncertainties.

\subsubsection{Nuclear Network and Decay Rates}
\label{sec: network and decay rates}

To accurately track energy generation and nucleosynthesis throughout the stellar
lifetime and the subsequent explosion, we employ a dedicated nuclear reaction
network coupled directly to the stellar structure equations. We utilize a custom
162-isotope network, referred to as \texttt{mesa162.net}, which is an extension
of the standard \texttt{mesa\_161.net}. The detailed list of isotopes included in 
the network is provided in Table~\ref{tab:isotopes}.

\begin{deluxetable}{lclc}[ht!]
\label{tab:isotopes}
\tablecaption{Isotopes Included in the Nuclear Reaction Network \label{tab:isotopes}}
\tablecolumns{4}
\tablewidth{0pt}
\tablehead{
    \colhead{Element} & \colhead{Mass Numbers ($A$)} & 
    \colhead{Element} & \colhead{Mass Numbers ($A$)}
}
\startdata
n  & 1                  & S  & 31--35 \\
H  & 1, 2               & Cl & 35--38 \\
He & 3, 4               & Ar & 35--40 \\
Li & 7                  & K  & 39--44 \\
Be & 7, 9, 10           & Ca & 39--46 \\
B  & 8, 10, 11          & Sc & 43--48 \\
C  & 12, 13             & Ti & 43--51 \\
N  & 13--15             & V  & 47--53 \\
O  & 14--18             & Cr & 47--57 \\
F  & 17--19             & Mn & 51--57 \\
Ne & 18--22             & Fe & 51--61 \\
Na & 21--24             & Co & 55--63 \\
Mg & 23--26             & Ni & 55--64 \\
Al & 25, 26\tablenotemark{a}, 27, 28 & Cu & 59--64 \\
Si & 27--31             & Zn & 60--64 \\
P  & 30--33              &            \\
\enddata
\tablenotetext{a}{The network explicitly treats the ground state ($^{26}\mathrm{Al}^g$) and the isomeric state ($^{26}\mathrm{Al}^m$) as separate species.}
\end{deluxetable}

A distinguishing feature of our network is the explicit treatment of the
\iso{Al}{26} branched decay. Rather than treating \iso{Al}{26} as a single
species, we separate it into its ground state $\iso{Al}{26}^g$ ($J^\pi=5^+$,
$t_{1/2}=7.17\times10^5$ yr) and its short-lived metastable isomeric state
$\iso{Al}{26}^m$ ($J^\pi=0^+$, $t_{1/2}=6.35$ s). These states are coupled via
temperature-dependent internal transitions, allowing our models to
self-consistently track the survival fraction of $\iso{Al}{26}^g$ under both
hydrostatic and explosive conditions without relying on post-processing. For the
iron-peak region, the network extends to \iso{Zn}{64}, specifically encompassing
isotopes from \iso{Fe}{51} to \iso{Fe}{61} and \iso{Co}{55} to \iso{Co}{63} to
ensure a robust representation of the pathways leading to \iso{Fe}{60}.

For computational efficiency during preliminary resolution sensitivity tests
(Section~\ref{subsubsec:resolution}), we utilize a smaller hardwired
network, \texttt{approx21\_cr60\_plus\_co56.net}, which has been extensively
validated for massive star energetics
\citep{
     1978ApJ...225.1021W,
     1988PhR...163...79W,
     2000ApJ...528..368H,
     2010ApJ...724..341H,
     2012ApJ...748...42C,
     2016ApJS..227...22F}.

\begin{figure}[htbp]
\centering
\includegraphics[width=0.48\textwidth]{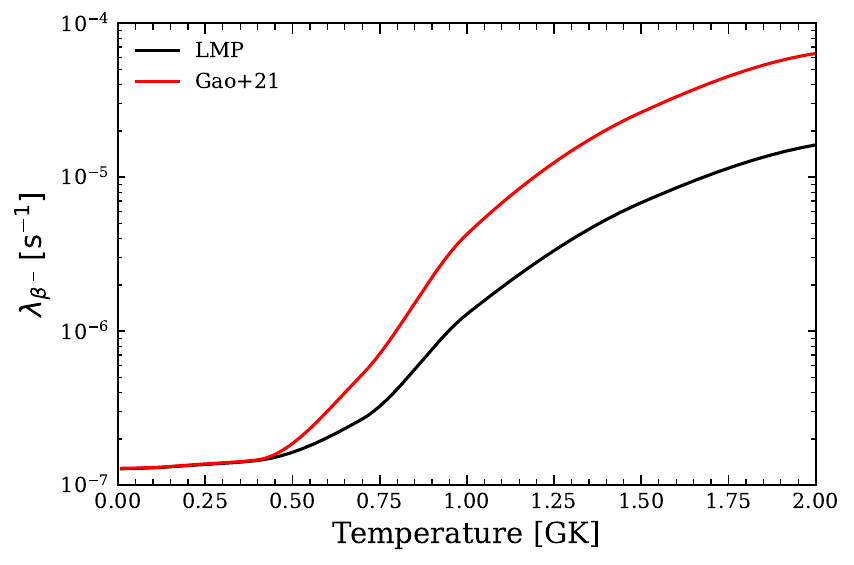}
\caption{The $\beta^-$ decay rate of \iso{Fe}{59} as a function of
temperature. The black line represents the widely used rates from
\citet{2001ADNDT..79....1L},
evaluated at a typical carbon-shell density
of $\rho Y_e \approx 10^5\,\mathrm{g\,cm^{-3}}$, while the red line shows the
new rate from \citet{2021PhRvL.126o2701G}. The temperature range is truncated at
2\,GK, as the $\beta^-$ decay channel becomes sub-dominant relative to
the \iso{Fe}{59}(p,n)\iso{Co}{59} reaction at higher temperatures.}
\label{fig:decay_rates}
\end{figure}

All reaction rates are drawn from \texttt{JINA REACLIB}
\citep[version 20171020;][]{2010ApJS..189..240C}
with the exception of the \iso{Fe}{59} $\beta^-$ decay rate. The \iso{Fe}{59}
$\beta^-$ decay rate provided in \texttt{JINA REACLIB} is taken from LMP, which
is widely adopted in stellar evolution and nucleosynthesis studies
\citep[e.g.,][]{2002RvMP...74.1015W, 2011ApJS..192....3P, 2018ApJS..237...13L}.
However, recent experimental work by \citet{2021PhRvL.126o2701G} indicates that
the decay rate is approximately three times higher than the LMP estimate at
temperatures relevant to C-shell burning ($T \approx 1.2$\,GK) as shown in
Figure \ref{fig:decay_rates}. Regarding the production channel, we adopt the
default \iso{Fe}{59}$(n, \gamma)$\iso{Fe}{60} rate from \texttt{JINA REACLIB}
for our entire model grid. Although recent experimental constraints on this
neutron capture cross-section have emerged \citep{2024NatCo..15.9608S}, we focus
specifically on isolating the impact of the updated $\beta^-$ decay rate in this
work. Given that the yield of \iso{Fe}{60} depends sensitively on the
competition between these two channels, the systematic uncertainties associated
with the neutron capture rate are further addressed in Section
\ref{subsubsection: nuclear uncertainties}.

\subsubsection{Numerical Resolution Strategy}
\label{subsubsec:resolution}
Reliable nucleosynthesis yields in massive stars require a fine-tuned balance
between spatial and temporal resolution, particularly to resolve the thin
burning shells and sharp composition gradients in advanced stages. We performed
a comprehensive resolution sensitivity analysis focusing on two key
\texttt{MESA} controls: the maximum fraction of a star's mass in a cell, 
\texttt{max\_dq} and the upper limit of the maximum allowed change in
the central density between timesteps,
\texttt{delta\_lgRho\_center\_limit}.

We adopt a two-stage resolution strategy based on convergence tests across our
mass range ($15, 25, 40\,M_{\odot}$). For the evolution from ZAMS to helium
depletion, we utilize a baseline resolution of \texttt{max\_dq} = $1.0 \times
10^{-3}$ and \texttt{delta\_lgRho\_center\_limit} = $1.5 \times 10^{-2}$. Taking
our $25\,M_{\odot}$ model as a representative case, this setup typically yields
a spatial grid of 2,000--3,000 mass shells, with time steps ranging from $\sim
10^{5}$\,yr during the main sequence to $\sim 10^{2}$\,yr approaching helium
depletion. This ensures that the helium core mass ($M_{\text{He}}$) converges
within a $5\%$ threshold. For the subsequent advanced burning stages until core
collapse, we switch to a significantly finer resolution with \texttt{max\_dq} =
$5.0 \times 10^{-4}$ and \texttt{delta\_lgRho\_center\_limit} = $5.0 \times
10^{-3}$ to capture the rapid structural changes and ensure a stable temperature
profile at the onset of collapse. For the $25\,M_{\odot}$ model, this enhanced
resolution results in a grid of 4,000--5,000 mass shells, with time steps
ranging from $\sim 10^{2}$\,yr to as low as $\sim 10^{-8}$\,yr (less than 1
second) to adaptively capture the rapid structural changes and convective
episodes in the final stages. While the $25$ and $40\,M_{\odot}$ models show
well-defined stability windows, we note that $15\,M_{\odot}$ models exhibit a
higher sensitivity to numerical choices, requiring a localized ``island of
stability" for consistent results. 

This rigorous control is essential because the \iso{Fe}{60} yield is extremely
sensitive to the thermal structure of the carbon-burning shells. Our tests
indicate that varying the resolution alone can induce yield fluctuations of up
to $60\%$, which could potentially obscure the $\sim 40\%$ physical suppression
expected from the updated \iso{Fe}{59} decay rate \citep{2021PhRvL.126o2701G}.
By prioritizing the convergence of the pre-supernova temperature profile, we
ensure that the suppressed \iso{Fe}{60} production observed in our grid is a
robust physical response to the nuclear physics update rather than a numerical
artifact. 

Detailed methodologies, convergence diagnostics, and the results of
our sensitivity grid are provided in Appendix~\ref{sec:appendix_resolution}.

\subsection{Core-collapse Supernovae} 
\label{sec:ccsn}

To simulate the hydrodynamic shock and explosive nucleosynthesis during the CCSN
event, we utilize the \texttt{CCSN} module in \texttt{MESA} (modeled after the
\texttt{example\_ccsn\_IIp} test suite) using a thermal bomb prescription as
described below. Initially, we define the mass cut at an entropy of $S =
4\,k_{\rm B}\,\text{baryon}^{-1}$ \citep{2010ApJ...724..341H,
2016ApJ...818..124E}. The material interior to this surface is excised from the
computational domain and assumed to form the compact remnant. To initiate the
explosion, we inject energy into a thin shell of $0.02\,M_\odot$, so that the
total energy of stars reaches $1 \times 10^{51}$\,erg
(\texttt{inject\_until\_reach\_model\_with\_total\_energy} = $1.0 \times
10^{51}$\,erg). This approach ensures that the resulting explosion carries a
prescribed net energy, accounting for the gravitational binding energy of the
progenitor's envelope. The energy is deposited over a characteristic timescale
of $t_{\rm inj} = 5$\,ms, launching a hydrodynamic shockwave that propagates
outward.

As the shockwave traverses the stellar layers, explosive nucleosynthesis
modifies the abundances of \iso{Al}{26} and \iso{Fe}{60}, primarily in the
carbon-burning (C) shell where post-shock temperatures exceed $2 \times 10^9$\,K
\citep{2006ApJ...647..483L}. In this study, we define the ``ejecta'' as mass
cells maintaining a positive total energy ($E_{\rm tot} = E_{\rm kin} + E_{\rm
th} + \Phi_{\rm grav} > 0$). The simulation is terminated when
the temperature of the innermost ejectable cell drops below $1.0 \times
10^8$\,K. At this stage, the charged-particle reactions responsible for the
production and destruction of \iso{Al}{26} and \iso{Fe}{60} have effectively
frozen out, ensuring the final yields remain constant.

\subsection{Galactic \iso{Fe}{60}/\iso{Al}{26} Flux Ratio}
\label{sec:imf_integration}

To evaluate the collective impact of the updated \iso{Fe}{59} decay rate at a
Galactic scale, we employ an integration framework to estimate the Galactic
$\gamma$-ray flux ratio, $I(\iso{Fe}{60})/I(\iso{Al}{26})$. Given that the
half-lives of \iso{Al}{26} ($t_{1/2} \approx 0.717$ Myr) and \iso{Fe}{60}
($t_{1/2} \approx 2.62$ Myr) are several orders of magnitude shorter than the
Galactic evolution timescale, we adopt the steady-state assumption where the
total production and decay rates are balanced
\citep[e.g.,][]{2006ApJ...647..483L, 2021PASA...38...62D}. Assuming both
isotopes originate from the same massive star populations, the averaged Galactic
$\gamma$-ray flux ratio can be directly related to the ratio of their integrated
Galactic yield rates:

\begin{equation}
\label{eq:flux_ratio}
     \frac{I(\iso{Fe}{60})}{I(\iso{Al}{26})} \approx 
     \frac{26}{60} \cdot 
     \frac{\dot{M}_{\text{Gal}}(\iso{Fe}{60})} 
     {\dot{M}_{\text{Gal}}(\iso{Al}{26})}
\end{equation}

The integrated Galactic yield rate for each isotope is calculated by convolving
our stellar yields with a normalized Initial Mass Function (IMF):

\begin{equation}
\label{eq:imf_int}
     \dot{M}_{\text{Gal}} \propto
     \int_{M_{\text{min}}}^{M_{\text{max}}} Y(m) \cdot m^{-(1+x)} dm 
\end{equation}

where $Y(m)$ is the total net mass of the isotope ejected by a star of initial
mass $m$ over its entire lifetime. For \iso{Fe}{60}, $Y(m)$ is derived from the
core-collapse supernova (CCSN) ejecta. For \iso{Al}{26}, $Y(m)$ represents the
cumulative sum of the time-integrated stellar wind contribution and the
explosive CCSN yield. In our calculations, the integration limits are set to
$M_{\text{min}}=14\,M_\odot$ and $M_{\text{max}}=80\,M_\odot$, consistent with
our \texttt{MESA} grid coverage. We adopt a baseline Salpeter IMF slope of $x =
1.35$ \citep{1955ApJ...121..161S} and further explore the sensitivity of the
flux ratio by varying $x$ from $0.5$ to $2.5$. This specific formulation allows
us to isolate the influence of the nuclear reaction rates from complex Galactic
transport effects, providing a direct measure of how the \iso{Fe}{59} decay rate
propagates from a single stellar interior to observable Galactic signals.

\section{Results} 
\label{sec:results}

We performed two parallel sets of stellar evolution and nucleosynthesis
calculations for models with initial masses spanning $14$ to $80\,M_{\odot}$.
Both grids utilize the extended \texttt{mesa\_162.net} nuclear network, as
detailed in Section~\ref{sec: network and decay rates}. The first grid,
hereafter the Default Rate (DR) series, employs the default \iso{Fe}{59}
$\beta^{-}$-decay rate from the LMP tables. The second grid, the New Rate (NR)
series, incorporates the updated experimental rate from
\citet{2021PhRvL.126o2701G}. All other nuclear inputs, physical and numerical
parameters remain identical between the two sets. This section presents the
direct numerical outputs of our simulations and is organized as follows.

We first describe the general evolutionary properties of our models, focusing on
the helium core masses ($M_{\text{He}}$) and central \iso{C}{12} mass fractions
at the end of core helium burning. We then present the yields for both
\iso{Al}{26} and \iso{Fe}{60}, and trace how the updated \iso{Fe}{59} decay rate
influences the nucleosynthetic yields of \iso{Fe}{60}, with the case pair of
$30\,M_{\odot}$ models. Finally, we present the resulting
\iso{Fe}{60}/\iso{Al}{26} flux ratios as a function of the IMF slope. 

\subsection{Basic Evolutionary Properties}
\label{sec:results_1}

Table \ref{tab:basic_evolution} summarizes the fundamental evolutionary
properties for both the DR and NR model grids. The table lists
the initial ($M_{\rm ini}$) and final ($M_{\rm fin}$) stellar masses, alongside
the structural parameters at helium depletion. We define helium depletion as the
stage when the central helium mass fraction drops below $X_{\rm
c}(^{4}\text{He}) = 10^{-4}$. At this stage, we determine the dimensions of the
helium core and the carbon-oxygen (CO) core based on chemical abundance
criteria. Specifically, the helium core boundary is defined as the outermost
coordinate where $X(^1\text{H})<1\times 10^{-4}$. For WR stars,
which have been stripped of their hydrogen envelopes via strong stellar winds,
the helium core mass is effectively equivalent to the total stellar mass.
Similarly, the CO core boundary is located where $X(^{4}\text{He})$ drops below
0.01 and either $X(^{12}\text{C})$ or $X(^{16}\text{O})$ exceeds 0.1.
Additionally, we report the central mass fraction of $^{12}\text{C}$ at
helium depletion.

Table \ref{tab:basic_evolution} also reports the properties characterizing the
CCSN explosions, including the initial mass cut ($M_{\rm cut}$), defined at
entropy $S = 4\,k_{\rm B}\,\text{baryon}^{-1}$, the fallback mass ($M_{\rm
fb}$), the final remnant mass ($M_{\rm rem}$), and the total ejecta mass
($M_{\rm ej}$). Finally, the relevant timescales are provided, covering the
lifetimes of hydrogen burning, helium burning, and the total stellar lifespan.

\begin{figure}[h]
    \centering
    \includegraphics[width=0.48\textwidth]{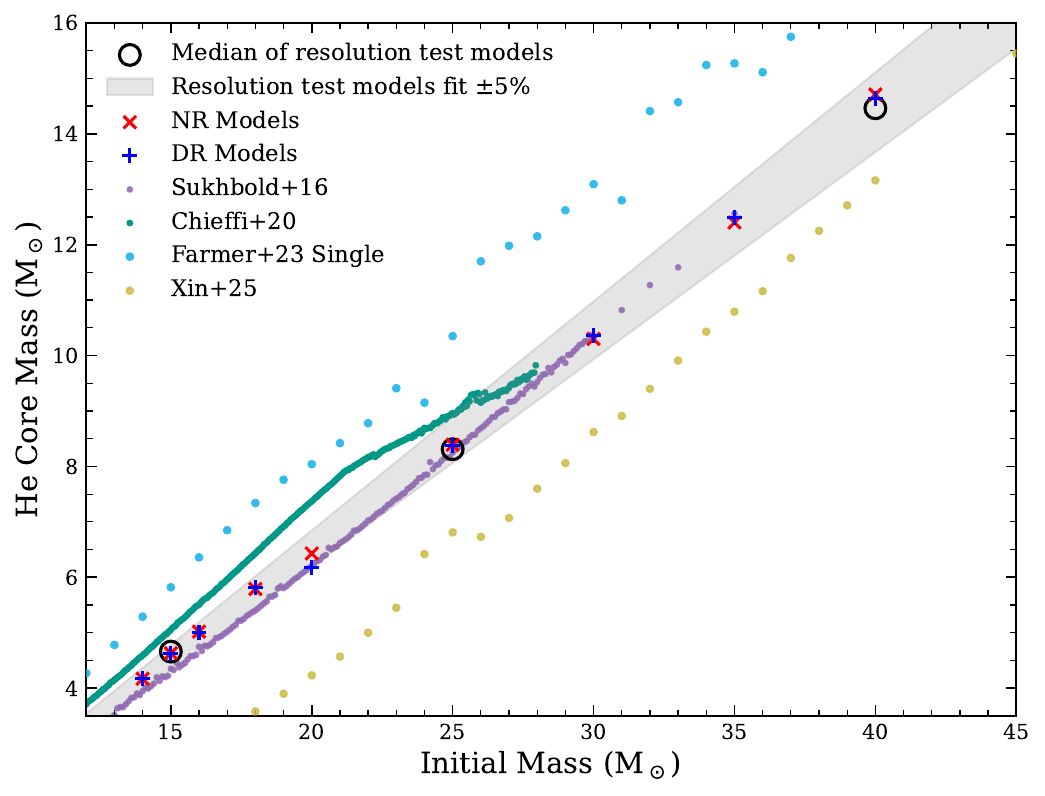}
     \caption{    Comparison of helium core masses ($M_{\text{He}}$) as a function of initial
    stellar mass ($M_{\text{ini}}$). The gray shaded region delineates the $\pm
    5\%$ convergence band derived from our resolution sensitivity analysis
    (Section \ref{subsubsec:resolution}). Our production models using both
    default (DR; blue pluses) and updated (NR; red crosses) \iso{Fe}{59} 
    decay rates are overlaid, showing no significant structural divergence. For
    benchmarking, we include data from 
    progenitor models of \citet{2016ApJ...821...38S},
    \citet{2020ApJ...890...43C},
    single-star models from \citet{2023ApJ...948..111F},
    and \citet{2025arXiv250211012X}. 
    }
     \label{fig:hecore_compare}
\end{figure}

The helium core mass ($M_{\text{He}}$) and central mass fraction of $^{12}\text{C}$
are two primary determinants of the advanced evolutionary stages 
and the final nucleosynthetic yields of massive stars
\citep{2020ApJ...890...43C, 2025arXiv250211012X}.
Figure~\ref{fig:hecore_compare} illustrates the relationship between
$M_{\text{He}}$ and the initial stellar mass ($M_{\text{ini}}$). We find that
$M_{\text{He}}$ exhibits a monotonic increase with $M_{\text{ini}}$ across the
entire range of $14$--$40\,M_{\odot}$. Notably, the $M_{\text{He}}$ values from
our DR and NR series are virtually identical, except the models with initial
mass of 20$\,M_{\odot}$, though the discrepancy remains within the established
numerical tolerance. Furthermore, both production series show excellent
agreement with the median values from our resolution sensitivity tests (see
Appendix~\ref{sec:appendix_resolution}). All production models fall within the
gray shaded region representing a $\pm 5\%$ deviation from the resolution
baseline, thereby validating the numerical robustness of our expanded
\texttt{mesa162.net} network. 

To ensure our models are consistent with the broader literature, we compare our
results against several grids calculated using different stellar evolution codes
: the \texttt{KEPLER} progenitor models from \citet{2016ApJ...821...38S} (S16);
the \texttt{FRANEC} models from \citet{2020ApJ...890...43C} (C20); and
\texttt{MESA} single-star models from \citet{2021A&A...645A...5S} (S21),
\citet{2023ApJ...948..111F}(F23), and \citet{2025arXiv250211012X} (X25).
Specifically, the $M_{\text{He}}$--$M_{\text{ini}}$ relation in our grid
demonstrates a high degree of conformity with the extensive progenitor sample
from S16. C20 and F23 yield higher core masses, while X25 shows a lower trend;
these differences primarily arise from differing treatments of convective
overshooting and mass loss prescriptions. Overall, this consistent alignment
provides a robust physical foundation for our subsequent yield analysis.

The central mass fraction of \iso{C}{12} at central helium depletion is a key
determinant of the subsequent shell-burning structures and nucleosynthesis
yields. \citet{2010ApJ...718..357T} demonstrated that the triple-$\alpha$ and
$\iso{C}{12}(\alpha, \gamma)\iso{O}{16}$ reaction rates influence the advanced
burning stages and final yields of \iso{Fe}{60} and \iso{Al}{26} in a complex
manner. In particular, the final yield of \iso{Fe}{60} is highly sensitive to
these rate uncertainties, potentially varying by a factor of $3\text{--}5$
within their $1\sigma$ uncertainty limits, whereas the \iso{Al}{26} yield is
less sensitive to these reaction rates. Consequently, the central mass fraction
of \iso{C}{12} at helium depletion serves as a critical benchmark for validating
the evolutionary baseline of our investigation into the nucleosynthesis of
\iso{Fe}{60}. 

\begin{figure}[h]
    \centering
    \includegraphics[width=0.48\textwidth]{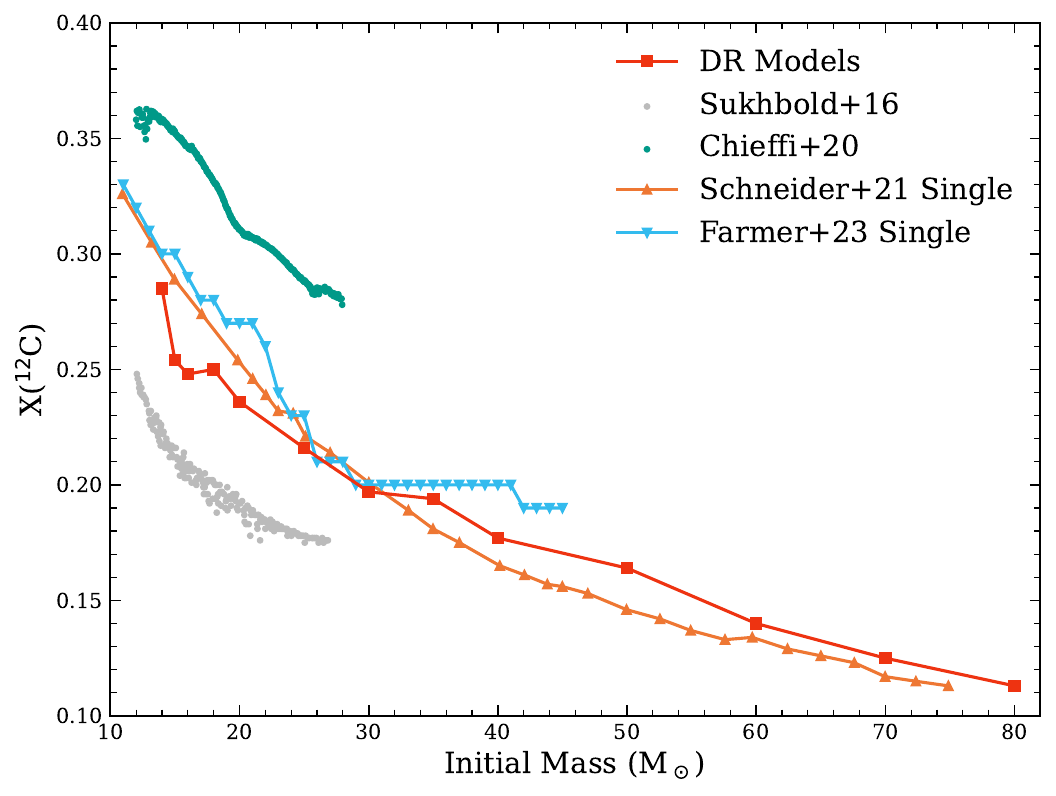}
     \caption{Comparison of \iso{C}{12} mass fractions at central helium
     depletion. Red solid line with squares: DR models; 
     grey dots: progenitor models from \citet{2016ApJ...821...38S}; 
     green dots: \citet{2020ApJ...890...43C}; 
     orange solid line with triangles: single-star models from 
     \citet{2021A&A...645A...5S}; 
     light blue solid line with triangles: single-star models from
     \citet{2023ApJ...948..111F}.
    }
     \label{fig:xc12_compare}
\end{figure}

Here we also compare our results against the reference grids.
Since the updated \iso{Fe}{59} decay rate does not physically impact the core
helium burning phase, the $X(\iso{C}{12})$ values between the DR and NR series
are virtually identical, subject only to minor numerical fluctuations (see Table
\ref{tab:basic_evolution}). For clarity, we only present the results of the DR
models in Figure \ref{fig:xc12_compare}. The overall trend of our models shows
excellent agreement with other \texttt{MESA}-based results (S21 and F23).
Notably, our carbon mass fractions, along with those from independent
\texttt{MESA} models, are positioned between the higher values predicted by the
\texttt{FRANEC} models (C20) and the lower values from the \texttt{KEPLER}
models (S16). This well-constrained $X(\iso{C}{12})\text{--}M_{\rm ini}$
relationship provides a reliable physical background for the subsequent shell
burning and s-process nucleosynthesis of \iso{Fe}{60}.

\begin{deluxetable*}{ccccccccccccc}
\label{tab:basic_evolution}
\tablewidth{0pt}
\tablecaption{Basic evolution properties of the stellar models} 
\tablehead{ \colhead{\iso{Fe}{59} decay rate} & \colhead{$M_\text{ini}$} &
\colhead{$M_\text{fin}$} & \colhead{$M_\text{He}$} & \colhead{$M_\text{CO}$} &
\colhead{$M_\text{cut}$} & \colhead{$M_\text{fb}$} & \colhead{$M_\text{rem}$} &
\colhead{$M_\text{ej}$} & \colhead{$X(^{12}\text{C})$} &
\colhead{$\tau_\text{H}$} & \colhead{$\tau_\text{He}$} &
\colhead{$\tau_\text{total}$}\\
\colhead{} & \colhead{(M$_\odot$)} & \colhead{(M$_\odot$)} &
\colhead{(M$_\odot$)} & \colhead{(M$_\odot$)} & \colhead{(M$_\odot$)} &
\colhead{(M$_\odot$)} & \colhead{(M$_\odot$)} & \colhead{(M$_\odot$)} &
\colhead{ } & \colhead{(Myr)} & \colhead{(Myr)} & \colhead{(Myr)} }
\startdata
{  } & 14 & 13.290 & 4.199  & 2.489  & 1.46 & 0.00  & 1.46  & 11.83 & 0.287 &
12.856 & 1.454 & 14.340  \\
{  } & 15 & 12.910 & 4.663  & 2.888  & 1.57 & 0.00  & 1.57  & 11.34 & 0.257 &
11.700 & 1.364 & 13.088  \\
{  } & 16 & 13.433 & 5.037  & 3.217  & 1.78 & 0.00  & 1.78  & 11.66 & 0.253 &
10.736 & 1.237 & 11.994  \\
{  } & 18 & 14.367 & 5.847  & 3.909  & 1.49 & 0.00  & 1.49  & 12.87 & 0.247 &
9.303  & 1.034 & 10.354  \\
{  } & 20 & 17.316 & 6.223  & 4.238  & 1.48 & 0.50  & 1.98  & 15.33 & 0.204 &
8.152  & 1.054 & 9.220   \\
{  } & 25 & 19.789 & 8.423  & 6.231  & 1.58 & 0.52  & 2.10  & 17.69 & 0.214 &
6.430  & 0.724 & 7.164   \\
{DR} & 30 & 22.098 & 10.412 & 8.058  & 2.25 & 0.72  & 2.97  & 19.13 & 0.189 &
5.586  & 0.512 & 6.105   \\
{  } & 35 & 23.310 & 12.551 & 10.003 & 1.97 & 2.97  & 4.94  & 18.37 & 0.190 &
4.882  & 0.529 & 5.417   \\
{  } & 40 & 25.745 & 14.705 & 11.964 & 2.07 & 0.46  & 2.53  & 23.21 & 0.182 &
4.454  & 0.481 & 4.940   \\
{  } & 50 & 33.245 & 19.011 & 15.891 & 1.72 & 0.95  & 2.66  & 30.58 & 0.175 &
3.895  & 0.415 & 4.314   \\
{  } & 60 & 24.136 & 24.181 & 20.350 & 2.66 & 0.00  & 2.65  & 21.49 & 0.171 &
3.535  & 0.373 & 3.912   \\
{  } & 70 & 32.927 & 30.608 & 26.945 & 2.19 & 22.80 & 24.99 & 7.94  & 0.122 &
3.284  & 0.359 & 3.646   \\
{  } & 80 & 34.649 & 34.980 & 31.404 & 2.06 & 30.29 & 32.35 & 2.30  & 0.116 &
3.097  & 0.341 & 3.441   \\
\hline
{  } & 14 & 13.290 & 4.198  & 2.493  & 1.43 & 0.00  & 1.43  & 11.87 & 0.285 &
12.856 & 1.455 & 14.341  \\
{  } & 15 & 12.895 & 4.663  & 2.895  & 1.64 & 0.00  & 1.64  & 11.26 & 0.254 &
11.700 & 1.370 & 13.094  \\
{  } & 16 & 13.407 & 5.053  & 3.224  & 1.62 & 0.00  & 1.62  & 11.79 & 0.248 &
10.736 & 1.239 & 11.996  \\
{  } & 18 & 14.425 & 5.830  & 3.890  & 1.50 & 0.00  & 1.50  & 12.93 & 0.250 &
9.303  & 1.033 & 10.352  \\
{  } & 20 & 17.351 & 6.471  & 4.475  & 1.54 & 0.00  & 1.54  & 15.81 & 0.236 &
8.148  & 0.941 & 9.103   \\
{  } & 25 & 19.608 & 8.435  & 6.241  & 1.63 & 0.00  & 1.63  & 17.98 & 0.216 &
6.430  & 0.723 & 7.163   \\
{NR} & 30 & 22.645 & 10.354 & 7.962  & 1.86 & 0.79  & 2.65  & 20.00 & 0.197 &
5.491  & 0.605 & 6.104   \\
{  } & 35 & 24.206 & 12.458 & 9.871  & 2.50 & 2.47  & 4.97  & 19.24 & 0.194 &
4.967  & 0.445 & 5.418   \\
{  } & 40 & 24.972 & 14.783 & 12.056 & 1.68 & 0.58  & 2.26  & 22.71 & 0.177 &
4.454  & 0.482 & 4.941   \\
{  } & 50 & 30.904 & 19.547 & 16.448 & 2.53 & 10.45 & 12.98 & 17.93 & 0.164 &
3.893  & 0.418 & 4.315   \\
{  } & 60 & 31.621 & 25.045 & 21.666 & 2.05 & 18.88 & 20.93 & 10.69 & 0.140 &
3.535  & 0.380 & 3.919   \\
{  } & 70 & 32.506 & 30.580 & 27.011 & 1.75 & 20.92 & 22.66 & 9.84  & 0.125 &
3.284  & 0.357 & 3.644   \\
{  } & 80 & 38.048 & 35.742 & 31.776 & 2.08 & 27.28 & 29.36 & 8.68  & 0.113 &
3.097  & 0.341 & 3.441   \\
\enddata
\tablecomments{All masses are given in solar units ($M_\odot$) and time in Myr.
$M_\text{ini}$ and $M_\text{fin}$ are the initial and final stellar masses.
$M_\text{He}$ and $M_\text{CO}$ denote the He-core and CO-core masses at helium
depletion. $M_\text{cut}$ is the initial mass cut at entropy $S = 4\,k_{\rm
B}\,\text{baryon}^{-1}$. $M_\text{fb}$ and $M_\text{rem}$ represent the fallback
mass and final remnant mass, respectively, while $M_\text{ej}$ is the total
ejecta mass. $X_{\rm c}(^{12}\text{C})$ is the central carbon mass fraction at
helium depletion.}
\end{deluxetable*}

\subsection{Nucleosynthetic Yields of \iso{Al}{26} and \iso{Fe}{60}}
\label{sec:results_2}

The nucleosynthetic yields for \iso{Al}{26} and \iso{Fe}{60} are detailed in
Table \ref{tab:yields}. For each model, we report the final total ejected mass
($M_{\rm tot}$) of each isotope, along with the contributions from three
distinct sources: (1) stellar winds (relevant only for \iso{Al}{26}), (2) the
hydrostatic shell contribution before the explosion, and (3) the net
contribution from explosive nucleosynthesis and fallback during the CCSN event. 

The hydrostatic shell contribution is computed by integrating the isotopic
abundances from the initial mass cut to the surface using the
pre-supernova profiles. The net contribution from CCSN events is calculated by
taking the difference between the final ejected mass and the pre-supernova
inventory. This term accounts for the complex interplay between shock-induced
synthesis and dynamic fallback. While the shock wave can synthesize new isotopes
(or destroy existing ones via photodisintegration), a portion of the
pre-existing material may fail to escape and instead fall back onto the compact
remnant. Consequently, a positive sign indicates net production during the
explosion, while a negative sign implies a net loss of the pre-existing
isotopes, either due to nuclear destruction or sequestration via fallback.

\begin{deluxetable*}{cccccccccccc}
\label{tab:yields}
\tablewidth{0pt}
\tablecaption{\iso{Al}{26} and \iso{Fe}{60} yields} \tablehead{
\colhead{\iso{Fe}{59} decay rate} & \colhead{$M_\text{ini}$} & & 
\multicolumn{4}{c}{\iso{Al}{26} Yields} & & 
\multicolumn{3}{c}{\iso{Fe}{60} Yields} \\
\cline{4-7}
\cline{9-11}
\colhead{} & \colhead{} & & 
\colhead{$M_\text{tot}$} & \colhead{$M_\text{wind}$} & \colhead{$M_\text{sh}$} &
\colhead{$M_\text{exp}$} & & 
\colhead{$M_\text{tot}$} & \colhead{$M_\text{sh}$} & \colhead{$M_\text{exp}$}\\
\colhead{} & \colhead{(M$_\odot$)} & 
& \colhead{(M$_\odot$)} & \colhead{(M$_\odot$)} & \colhead{(M$_\odot$)} &
\colhead{(M$_\odot$)} & & 
\colhead{(M$_\odot$)} & \colhead{(M$_\odot$)} & \colhead{(M$_\odot$)} }
\startdata
{  } & 14 & & 3.99(-05) & 4.37(-09) & 2.24(-05) & $+1.75(-05)$ & & 3.05(-05) & 8.82(-06) & $+2.17(-05)$ \\
{  } & 15 & & 2.23(-05) & 1.42(-09) & 1.98(-05) & $+2.47(-06)$ & & 5.47(-05) & 2.24(-05) & $+3.23(-05)$ \\
{  } & 16 & & 4.70(-05) & 2.40(-09) & 7.13(-05) & $-2.43(-05)$ & & 2.68(-05) & 4.30(-05) & $-1.62(-05)$ \\
{  } & 18 & & 3.13(-05) & 1.04(-08) & 1.30(-05) & $+1.83(-05)$ & & 1.60(-05) & 1.76(-05) & $-1.58(-06)$ \\
{  } & 20 & & 5.47(-05) & 1.42(-07) & 2.86(-05) & $+2.60(-05)$ & & 3.75(-05) & 5.41(-05) & $-1.66(-05)$ \\
{  } & 25 & & 8.79(-05) & 5.48(-07) & 4.44(-05) & $+4.30(-05)$ & & 5.21(-05) & 4.37(-05) & $+8.43(-06)$ \\
{DR} & 30 & & 1.90(-04) & 1.62(-06) & 2.42(-04) & $-5.35(-05)$ & & 1.20(-04) & 1.41(-04) & $-2.07(-05)$ \\
{  } & 35 & & 7.88(-05) & 3.03(-06) & 9.73(-05) & $-2.15(-05)$ & & 1.66(-04) & 1.74(-04) & $-7.32(-06)$ \\
{  } & 40 & & 2.27(-04) & 4.46(-06) & 2.25(-04) & $-2.72(-06)$ & & 2.26(-04) & 2.34(-04) & $-8.17(-06)$ \\
{  } & 50 & & 1.57(-04) & 7.18(-06) & 1.26(-04) & $+2.36(-05)$ & & 1.64(-04) & 1.64(-04) & $+1.43(-07)$ \\
{  } & 60 & & 4.15(-04) & 4.68(-05) & 5.34(-04) & $-1.66(-04)$ & & 2.41(-04) & 2.64(-04) & $-2.23(-05)$ \\
{  } & 70 & & 4.92(-04) & 5.37(-05) & 1.75(-04) & $+2.63(-04)$ & & 6.18(-05) & 1.47(-04) & $-8.48(-05)$ \\
{  } & 80 & & 1.07(-04) & 9.10(-05) & 8.69(-05) & $-7.14(-05)$ & & 1.58(-05) & 1.15(-04) & $-9.90(-05)$ \\
\hline
{  } & 14 & & 2.07(-05) & 4.38(-09) & 1.78(-05) & $+2.89(-06)$ & & 3.22(-06) & 3.83(-06) & $-6.09(-07)$ \\
{  } & 15 & & 1.21(-05) & 1.16(-09) & 1.37(-05) & $-1.65(-06)$ & & 4.55(-05) & 1.87(-05) & $+2.69(-05)$ \\
{  } & 16 & & 2.96(-05) & 2.55(-09) & 2.06(-05) & $+8.95(-06)$ & & 2.07(-05) & 2.59(-06) & $+1.81(-05)$ \\
{  } & 18 & & 3.43(-05) & 8.88(-09) & 1.24(-05) & $+2.19(-05)$ & & 9.92(-06) & 1.07(-05) & $-7.58(-07)$ \\
{  } & 20 & & 9.29(-05) & 8.57(-08) & 8.62(-05) & $+6.64(-06)$ & & 6.68(-06) & 7.20(-06) & $-5.14(-07)$ \\
{  } & 25 & & 7.74(-05) & 5.26(-07) & 3.26(-05) & $+4.43(-05)$ & & 4.46(-05) & 4.65(-05) & $-1.85(-06)$ \\
{NR} & 30 & & 1.20(-04) & 1.20(-06) & 8.16(-05) & $+3.75(-05)$ & & 9.33(-05) & 9.89(-05) & $-5.59(-06)$ \\
{  } & 35 & & 2.62(-04) & 2.27(-06) & 4.00(-04) & $-1.40(-04)$ & & 7.77(-05) & 9.34(-05) & $-1.57(-05)$ \\
{  } & 40 & & 9.14(-05) & 5.23(-06) & 5.84(-05) & $+2.78(-05)$ & & 5.18(-05) & 5.16(-05) & $+1.78(-07)$ \\
{  } & 50 & & 1.45(-04) & 8.52(-06) & 1.97(-04) & $-6.06(-05)$ & & 1.50(-04) & 1.86(-04) & $-3.55(-05)$ \\
{  } & 60 & & 5.65(-04) & 2.68(-05) & 2.35(-04) & $+3.03(-04)$ & & 5.31(-05) & 1.32(-04) & $-7.90(-05)$ \\
{  } & 70 & & 1.19(-04) & 5.58(-05) & 1.23(-04) & $-6.02(-05)$ & & 6.86(-05) & 1.21(-04) & $-5.26(-05)$ \\
{  } & 80 & & 2.01(-04) & 7.27(-05) & 6.20(-05) & $+6.58(-05)$ & & 2.99(-05) & 7.88(-05) & $-4.89(-05)$ \\
\enddata
\tablecomments{All yields are given in solar masses ($M_\odot$). Parentheses
denote powers of ten, where $A(B)$ signifies $A \times 10^{B}$ (e.g., $2.06(-04)
= 2.06 \times 10^{-4}$). $M_{\text{tot}}$ is the sum of wind, shell, and
explosive components. For $M_{\text{exp}}$, a positive value indicates net
production, while a negative value implies net decrease after explosion.}
\end{deluxetable*}

\begin{figure*}[htbp]
     \centering
     \includegraphics[width=1.0\textwidth]{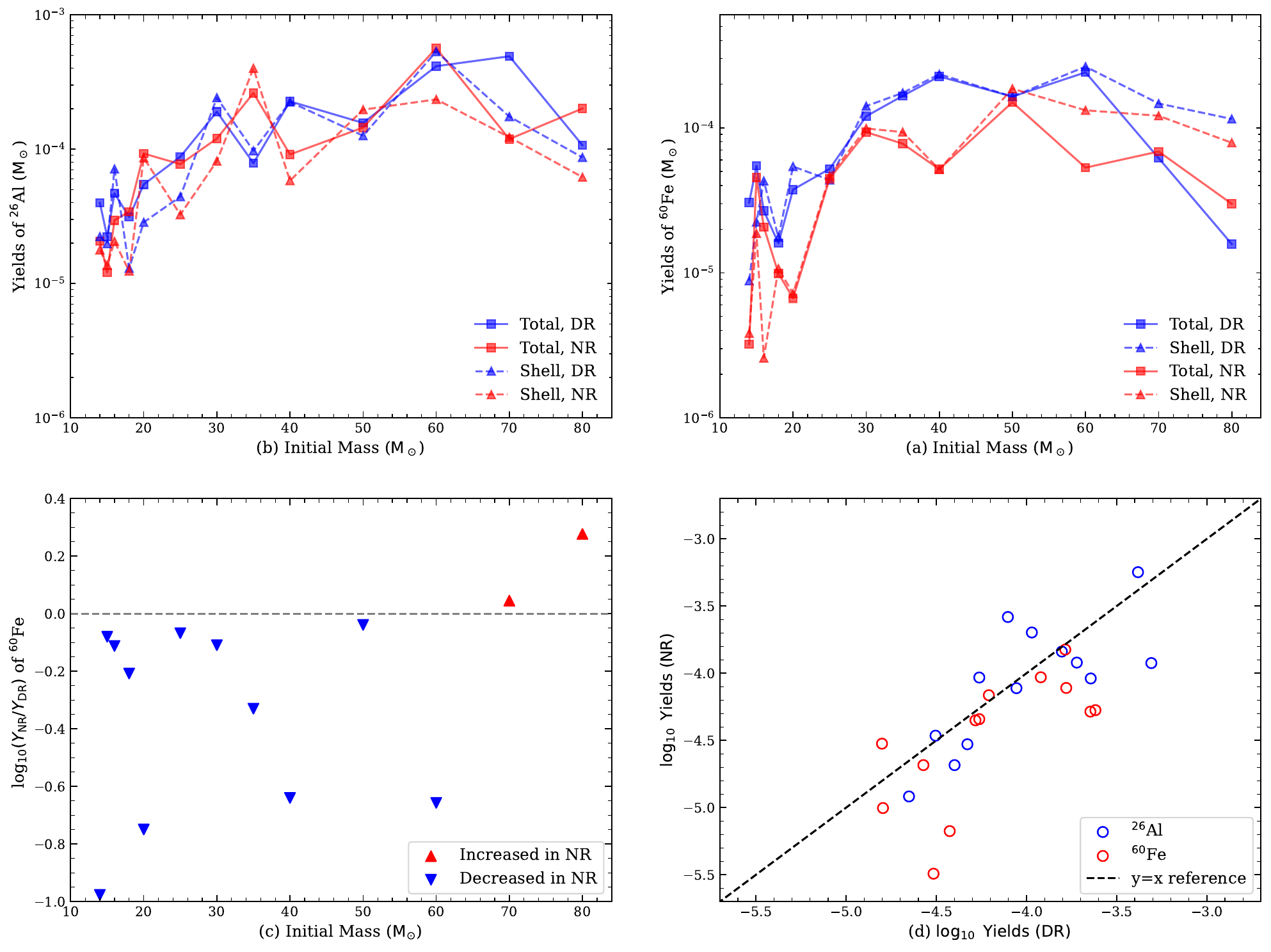}
     \caption{
     Nucleosynthesis yields of \iso{Al}{26} and \iso{Fe}{60} as a
     function of initial mass ($14\text{--}80\,M_{\odot}$). (a, b) Comparison of
     yields between models using the default rate (DR, blue lines) and the
     updated rate (NR, red lines). Solid lines represent total yields, while
     dashed lines indicate pre-supernova hydrostatic shell yields. 
     (c) The logarithmic ratio of \iso{Fe}{60} total yields
     ($\log_{10}(Y_{\text{NR}}/Y_{\text{DR}})$), highlighting the substantial
     suppression of production in the NR models. (d) Direct comparison of yields
     relative to the identity line ($y=x$). Note the distinct response to the
     rate update: while \iso{Fe}{60} (squares) exhibits a systematic reduction,
     \iso{Al}{26} (circles) shows only random fluctuations.
     }
     \label{fig:yields}
\end{figure*}

We first examine the yields of \iso{Al}{26} as a control case to quantify
the intrinsic numerical fluctuations inherent in our 1D stellar models. Since
the production channels of \iso{Al}{26} are physically decoupled from the
iron-peak weak interactions, its yields should, in principle, remain unaffected
by the updated $^{59}\text{Fe}$ decay rate. As shown in Figures
\ref{fig:yields}(a) and (d), the \iso{Al}{26} yields remain robustly
consistent across both the Default Rate (DR) and New Rate (NR) grids. To
formally validate this, we calculated the logarithmic yield differences, $\Delta
\log_{10} Y = \log_{10}(Y_{\mathrm{NR}}) - \log_{10}(Y_{\mathrm{DR}})$. A
one-sample $t$-test on these differences yields $p = 0.457$, indicating that we
cannot reject the null hypothesis of zero expected change. This confirms that
the variations in \iso{Al}{26} exhibit no systematic physical offset and are
driven purely by stochastic numerical noise. Consequently, we adopt the standard
deviation of these fluctuations ($\sigma_{\mathrm{Al}} \approx 0.26$ dex) as the
baseline numerical noise floor for our subsequent analysis.

In stark contrast, the nucleosynthesis of \iso{Fe}{60} is profoundly
impacted by the updated nuclear physics input. Figures \ref{fig:yields}(b) and
(c) reveal a systematic suppression of \iso{Fe}{60} yields in the NR models
(red lines) compared to the DR models (blue lines). When analyzing the full mass
range, the mean physical signal of \iso{Fe}{60} reduction is partially
obscured by background noise ($\text{SNR} \approx 0.95$), primarily due to the
extreme massive end of our grid ($M_{\text{ini}} \ge 60\,M_{\odot}$). In these
cases, both model series exhibit a sharp decline in yields due to substantial
fallback during the CCSN explosion. In this fallback-dominated
regime, the final yields are heavily dictated by explosion dynamics rather than
nucleosynthesis processes. By focusing on the models less affected by
fallback, the physical signal of \iso{Fe}{60} suppression
strengthens to a mean difference of $\Delta \log_{10} Y(\mathrm{Fe}) \approx
-0.33$ dex, resulting in an improved signal-to-noise ratio of $\text{SNR} =
1.27$. This demonstrates that the suppression of \iso{Fe}{60} successfully
penetrates the $1\sigma$ numerical noise floor of the models.

To statistically secure the significance of this suppression, we performed a
paired $t$-test on the logarithmic yields of \iso{Fe}{60} across the entire
grid. The test yields a highly significant result ($t = -2.768$, $p = 0.017$),
formally confirming that the reduction is a robust physical consequence of the
modified $^{59}\text{Fe}$ decay rate rather than a numerical artifact.
Quantitatively, this modification leads to an average systematic reduction of
$0.28$ dex ($\sim 48\%$) in the total \iso{Fe}{60} yields, with the effect being
more pronounced ($0.33$ dex, or $\sim 53\%$) when excluding fallback-dominated
cases. 
\rev{This systematic offset suggests that the enhanced $\beta^-$ decay of
\iso{Fe}{59} acts as a primary mechanism for suppressing \iso{Fe}{60} production
during convective shell burning. However, we note that the overall reduction in
the \iso{Fe}{60} yield could be more modest than indicated by our current models
when considering competing effects, such as the recently measured enhancement in
the \iso{Fe}{59}($n, \gamma$)\iso{Fe}{60} capture rate
\citep{2024NatCo..15.9608S}. The net impact on the Galactic
\iso{Fe}{60}/\iso{Al}{26} ratio thus remains subject to future multi-channel
sensitivity tests that incorporate all updated nuclear data.}

\subsection{Physics of the Suppression: The $30\,M_{\odot}$ Case}
\label{sec:micro_physics}

To elucidate the physical mechanism driving this suppression, we specifically
examine the internal evolution of the $30\,M_\odot$ model. We select this mass
as a representative case because it effectively captures the interplay between
the weak interaction rates and stellar structure without the complicating
effects of fallback dominance found in higher masses.

As detailed in previous studies \citep{2006ApJ...647..483L,2021PASA...38...62D},
\iso{Fe}{60} production is governed by the neutron capture sequences during the
He and C burning phases, especially within the convective shells. During He
burning, temperatures typically remain below $0.5$\,GK. In this regime, both the
LMP and the updated decay rates are consistently low (see Figure
\ref{fig:decay_rates}), implying minimal impact from the new nuclear data.
However, at the higher temperatures characteristic of C-shell burning ($T
\gtrsim 1.0$\,GK), the updated rate becomes significantly faster than the LMP
prediction, creating a potent "leak" in the s-process path that suppresses the
final \iso{Fe}{60} yield.

\subsubsection{Temporal Evolution}
\begin{figure}[htbp]
     \centering
     \includegraphics[width=0.48\textwidth]{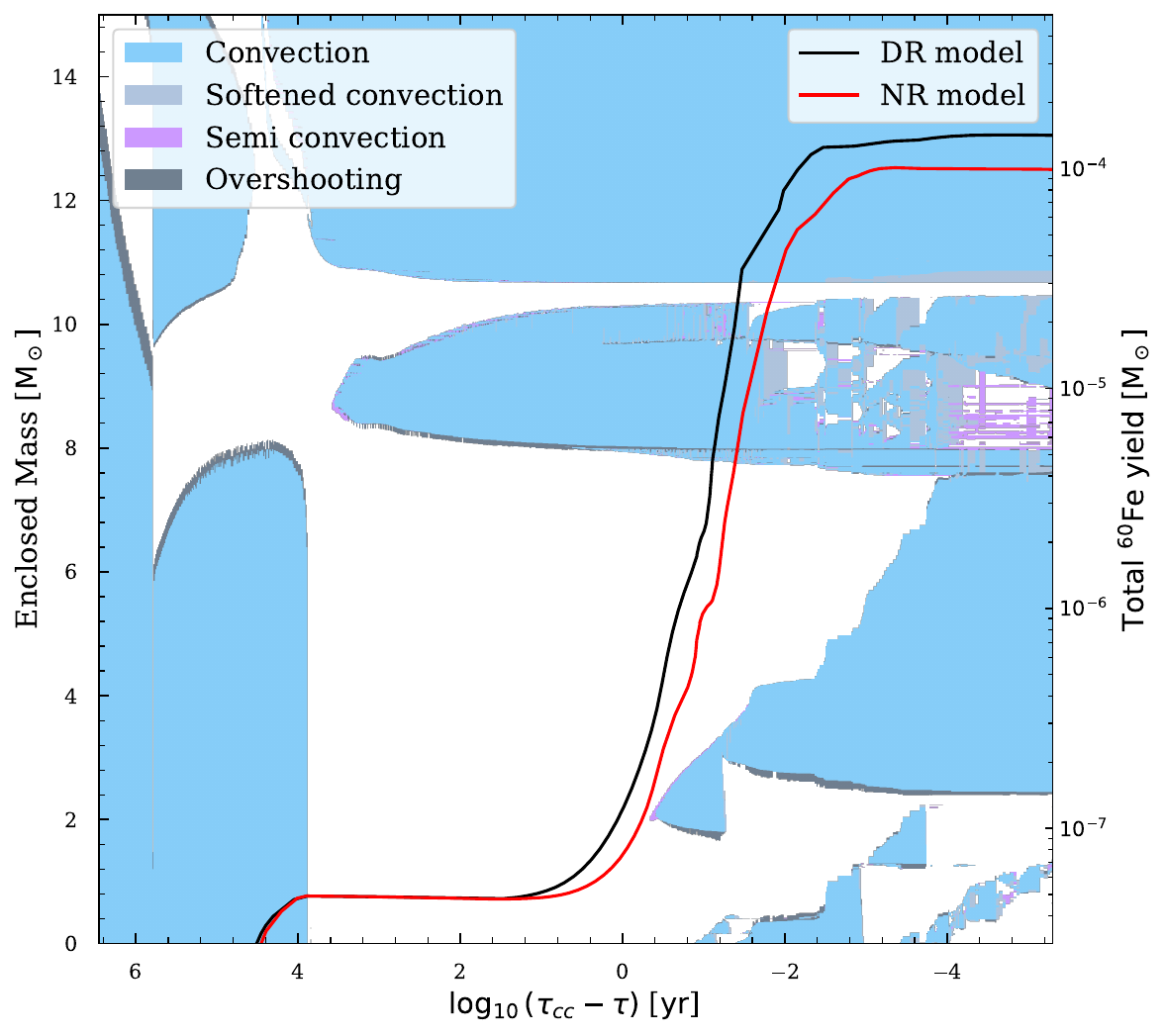}

     \caption{Evolutionary history of the $30\,M_\odot$ model. The background
     Kippenhahn diagram (left axis) illustrates the internal structure
     evolution, showing convective (light blue), semi-convective (purple), and
     overshooting (dark gray) zones as a function of the logarithmic time
     remaining until core collapse ($\tau_{cc} - \tau$). The overlaid solid
     lines (right axis) track the cumulative \iso{Fe}{60} mass yield. The black
     line represents the Default Rate (DR) model, while the red line indicates
     the New Rate (NR) model. Note that the deviation in \iso{Fe}{60} production
     between the two models becomes prominent during the late evolutionary
     stages (e.g., carbon burning).}

     \label{fig:kipp_yields_trace_30}
\end{figure}

Figure \ref{fig:kipp_yields_trace_30} illustrates this mechanism by tracking the
cumulative \iso{Fe}{60} mass alongside the internal convective history. As shown
by the solid lines, the accumulation tracks for the DR (black) and NR (red)
models are identical during the early evolutionary phases ($\log( \tau_{\rm cc}
- t) \gtrsim 1$). This confirms that during helium burning, the difference in
decay rates is negligible, resulting in identical nucleosynthetic paths.

A significant divergence emerges during the advanced stages, concomitant with
the onset of convective carbon shell burning. In both models, the total
\iso{Fe}{60} yield increases rapidly as the carbon shell develops, driven by the
activation of the \iso{Ne}{22}$(\alpha, n)$\iso{Mg}{25} neutron source. However,
the NR model exhibits a markedly slower growth rate due to the enhanced decay of
the precursor \iso{Fe}{59}. As the fuel and neutron source are consumed, the
total \iso{Fe}{60} yield in both models reaches a plateau prior to core silicon
burning. Ultimately, the enhanced decay rate in the NR model leads to a final
reduction of approximately $30\%$ compared to the DR model.

\subsubsection{Spatial Structure and Explosive Modification}

\begin{figure*}[htbp]
     \centering
     \includegraphics[width=1.0\textwidth]{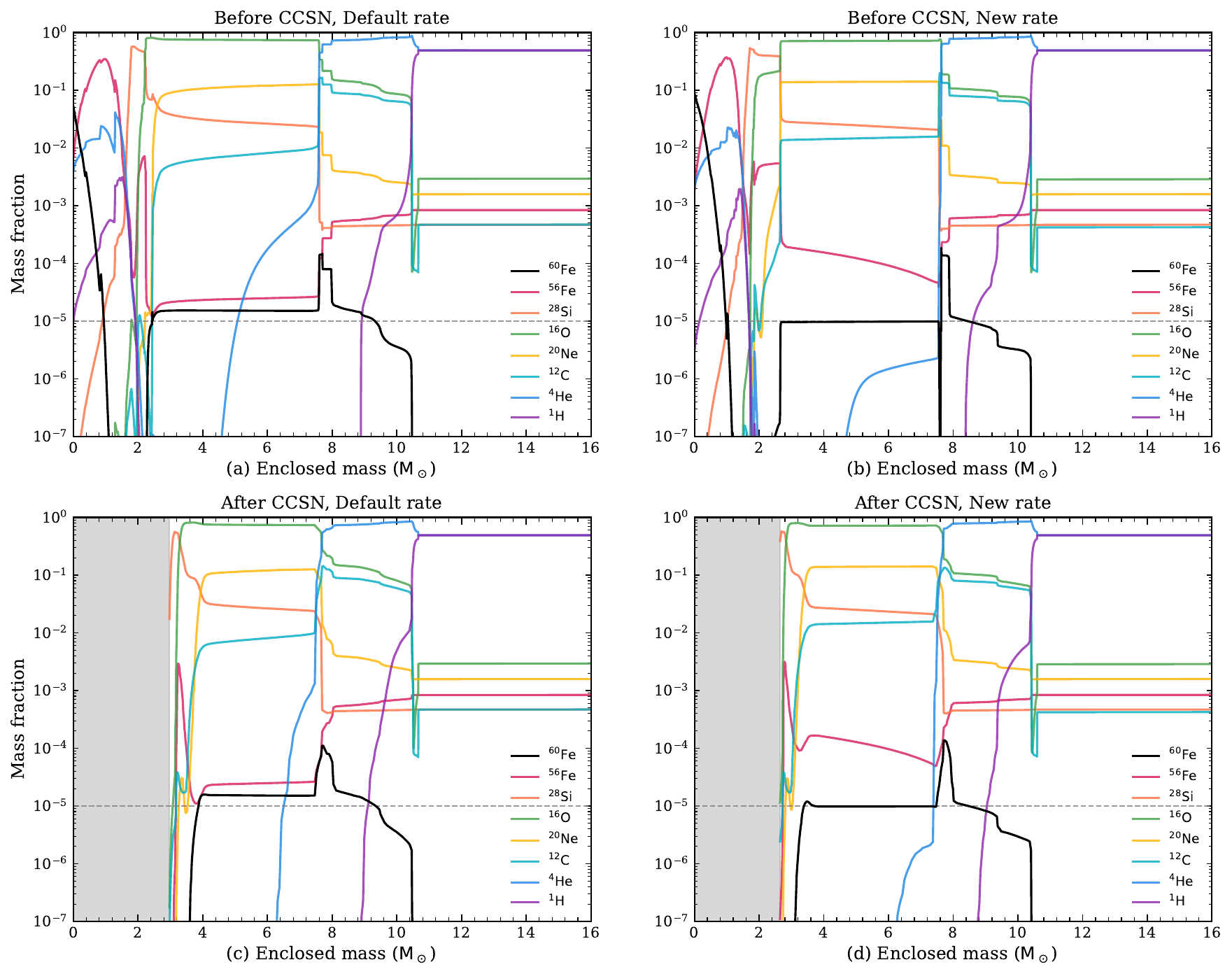}
     \caption{Internal mass fraction profiles of key isotopes (\iso{H}{1},
     \iso{He}{4}, \iso{C}{12}, \iso{O}{16}, \iso{Ne}{20}, \iso{Si}{28},
     \iso{Fe}{56}, and \iso{Fe}{60}) for the $30\,M_\odot$ models. The left
     panels (a, c) correspond to the Default Rate (DR) case, while the right
     panels (b, d) represent the New Rate (NR) case. Top panels show the
     pre-supernova stage, and bottom panels display the post-explosion
     distribution. The \iso{Fe}{60} abundance is highlighted by the solid black
     line, with a horizontal dashed gray line at $X = 10^{-5}$ provided for
     visual reference. The gray shaded regions in the bottom panels denote the
     mass enclosed within the compact remnant and the fallback material ($M <
     M_{\rm cut} + M_{\rm fb}$), which is excluded from the final ejecta.}
     \label{fig:mass_frac_compare_30}
\end{figure*}

The impact of the updated \iso{Fe}{59} decay rate on the spatial distribution of
\iso{Fe}{60} is presented in Figure \ref{fig:mass_frac_compare_30}, which
compares the internal mass fraction profiles before (top panels) and after
(bottom panels) the supernova explosion. The top panels reveal that while the
bulk shell structure remains robust against the rate update, the \iso{Fe}{60}
abundance is significantly altered. Specifically, in the DR model (top left),
the \iso{Fe}{60} mass fraction remains above $10^{-5}$ (grey dashed line) across
a wide range of the O/Ne and He shells, with a distinct peak at the He/O shell
boundary. In contrast, the NR model exhibits systematically lower abundances
throughout these layers, directly reflecting the suppressed hydrostatic
accumulation discussed above.

To further elucidate the spatial origin of this suppression, we decomposed the
total pre-supernova \iso{Fe}{60} yields ($M_{\text{sh}}$ of \iso{Fe}{60} Yields
in Table \ref{tab:yields}) into contributions from the C-burning shell and the
He-burning shell. Across our entire model grid, the He shell contributes an
average of $\sim\,26\%$ to the total \iso{Fe}{60} inventory, although this
fraction varies significantly with initial mass (ranging from nearly 0\% to
almost $80\%$). Statistical analysis reveals a clear site-specific sensitivity
to the updated nuclear data: the \iso{Fe}{60} produced in the C-burning
shell, where temperatures reach $T \approx 1.2$\, GK, exhibits a significant
systematic reduction of approximately 0.38 dex ($p = 0.017$, paired $t$-test) in
the NR grid. Conversely, the \iso{Fe}{60} abundance in the He-shell remains
virtually unaffected by the rate update, showing no statistically significant
offset. This disparity confirms that the global suppression of \iso{Fe}{60} is
primarily driven by the enhanced decay of \iso{Fe}{59} in the high temperature
environment of the C-shell, while the He-shell provides a stable component.

The bottom panels illustrate the modifications induced by the supernova shock.
We observe that the passage of the shock wave results in largely minor changes
to the global \iso{Fe}{60} profile, except at the very base of the ejecta. Here,
the high post-shock temperatures trigger the photodisintegration of \iso{Fe}{60}
into lighter species via $(\gamma,n)$ reactions, and the reaction \iso{Fe}{60}
$(p, n)$ \iso{Co}{60} can further deplete the abundance. Although a minor peak
corresponding to explosive production is visible, it is outweighed by the
destruction process. Consequently, for the $30\,M_\odot$ model, the net
contribution from the explosion is negative (as listed in Table
\ref{tab:yields}). This balance between explosive production and destruction
varies across the model grid and is inherently sensitive to the assumed
explosion properties. While we adopt a uniform explosion energy of
$10^{51}$\,erg (1\,foe) for all models, the peak post-shock temperatures, and
thus the extent of \iso{Fe}{60} destruction would scale with the explosion
energy. The systematic uncertainties associated with this simplified explosion
prescription and its impact on the final yields are further addressed in Section
\ref{subsec:discussion_3}.

\subsection{Galactic \iso{Fe}{60}/\iso{Al}{26} Flux Ratio}
\label{sec:results_flux}

\begin{figure}[htbp]
     \centering
     \includegraphics[width=0.48\textwidth]{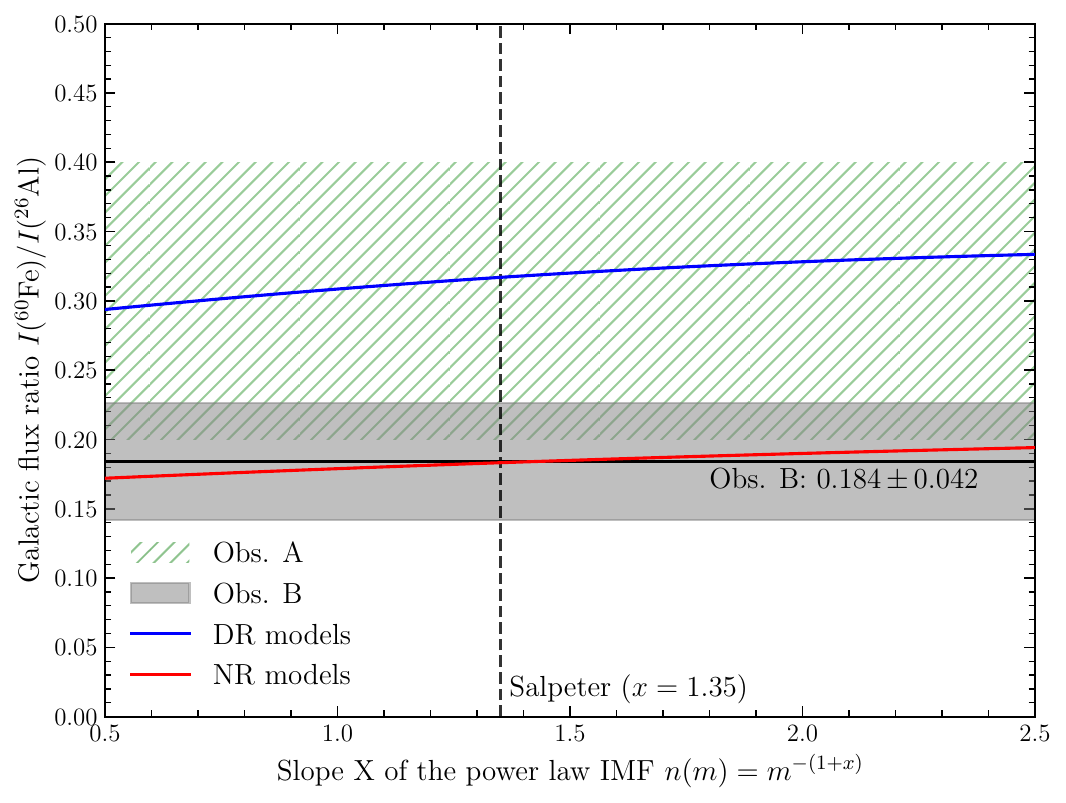}
    \caption{Predicted Galactic $\gamma$-ray flux ratio
     $I(\iso{Fe}{60})/I(\iso{Al}{26})$ as a function of the IMF slope parameter
     $x$. Results are shown for models with default \iso{Fe}{59} decay rates
     (DR, blue line) and updated decay rates (NR, red line). The gray shaded
     region (Obs. B) and green hatched area (Obs. A) denote the observational
     constraints from the exponential disk model ($0.184 \pm 0.042$) and various
     spatial morphology models, both derived from INTEGRAL data by
     \citet{2020ApJ...889..169W}. For reference, the horizontal solid line
     marks the observed central value of $0.184$, while the vertical dashed line
     marks the standard Salpeter IMF slope ($x = 1.35$).
    }
     \label{fig:flux_ratio}
\end{figure}

Based on the integration framework outlined in Section \ref{sec:methods}, we
predict the Galactic $\gamma$-ray flux ratio, $I(\iso{Fe}{60})/I(\iso{Al}{26})$,
utilizing our computed stellar yields. Figure \ref{fig:flux_ratio} compares
these predictions against the observational constraint of $0.184 \pm 0.042$,
derived from INTEGRAL observations using a parameterized spatial morphology
model \citep{2020ApJ...889..169W}. While we adopt this value as our primary
benchmark, we note that observational estimates can range from 0.2 to 0.4
depending on the assumed spatial distributions of the two isotopes. A more
detailed evaluation of these observational uncertainties is provided in Section
\ref{subsec:discussion_2}.

Assuming a standard Salpeter IMF slope of $x=1.35$, the DR models (blue line)
predict a flux ratio of $\sim 0.32$, which systematically overestimates the
observed value, consistent with previous theoretical studies. In contrast, the
NR models (red line), incorporating the updated reaction rate, yield a
significantly reduced ratio of $\sim 0.18$. This value is in excellent agreement
with the observational limits.

Notably, the predicted flux ratio exhibits a weak dependence on
the IMF slope parameter $x$. As shown in Figure \ref{fig:flux_ratio}, the ratio
remains relatively stable across a wide range of $x$ values ($0.5 \le x \le
2.5$), with variations of less than $\sim 0.05$. 

The weak dependence of the integrated flux ratio on the IMF slope $x$ can be
attributed to the yield distribution across our model grid. Specifically, the
$16\text{--}50\,M_\odot$ mass range represents the dominant source for Galactic
\iso{Fe}{60} and \iso{Al}{26}, consistently accounting for approximately $70\%$
of the total integrated flux regardless of the chosen $x$ value ($0.5 \le x \le
2.5$). Within this bulk population, the single-star \iso{Fe}{60}/\iso{Al}{26}
ratios remain relatively stable, effectively anchoring the Galactic average.
Furthermore, in IMF scenarios with small $x$, one might expect the most massive
stars to significantly shift the ratio with their high \iso{Fe}{60}/\iso{Al}{26}
yields. However, in our models, stars with $M_{\text{ini}} \ge 60\,M_\odot$
undergo substantial fallback during the supernova explosion with the applied
explosion energy. This process traps a large fraction of the \iso{Fe}{60}
produced in the inner layers, preventing these very massive stars from
exhibiting the high \iso{Fe}{60}/\iso{Al}{26} ratios that would otherwise
inflate the integrated result. Consequently, the Galactic flux ratio is robustly
determined by the dominant intermediate mass population, making it largely
insensitive to the specific choice of IMF parameters.

However, we caution against overinterpreting the apparent alignment between our
NR prediction ($\approx 0.18$) and the observational central value ($0.184$). Given
the systematic uncertainties remaining in both stellar modeling and $\gamma$-ray
data analysis, this excellent agreement may be partly fortuitous. Consequently,
rather than serving as a final confirmation, this consistency and the IMF
insensitivity underscore the need for a critical assessment of the underlying
uncertainties, which we address in detail in Section \ref{sec:discussion}.

\section{Discussion}
\label{sec:discussion}

Although our NR models incorporating the updated \iso{Fe}{59} decay rate
demonstrate the significant suppression of \iso{Fe}{60} production and yield a
galactic flux ratio consistent with observations, it is crucial to interpret
these results within the context of model dependencies. In this section, we
discuss the mass-dependent trends, compare our integrated results with
observational constraints, and address the inherent systematic uncertainties.

\subsection{Mass Dependence of the \iso{Fe}{60} Yields and the Role of Fallback}
\label{subsec:discussion_1}

The distinctive mass-dependent features of \iso{Fe}{60} production are most
prominent in the lower mass range ($M_{\rm ini} < 50\,M_\odot$). In this regime,
the \iso{Fe}{60} yields show a general upward trend with initial mass. However,
detailed structure exists: the NR grid reveals a double-peak structure with
maxima at $\sim 15\,M_\odot$ and $\sim 30\,M_\odot$, whereas the DR grid
displays a single peak at $\sim 15\,M_\odot$. Similar monotonic behavior with
local variations has been reported in prior studies \citep{2006ApJ...647..483L},
although the exact peak locations differ. These shifts in peak positions likely
reflect differences in underlying stellar physics assumptions (e.g., mass-loss
rates and convective overshoot efficiencies) rather than the nuclear input
alone. As analyzed in Section \ref{sec:results_flux}, this intermediate-mass
range serves as the primary engine for Galactic enrichment, accounting for
$\sim\,70\%$ of the integrated flux ratio.

For high-mass stars, particularly those with $M_{\rm ini} > 50\,M_\odot$, a
noticeable trend is the sharp decline in net \iso{Fe}{60} yields. As shown in
Figure \ref{fig:yields}(a), the final total yields (solid lines) in this regime
tend to be lower than the pre-supernova hydrostatic contributions (dashed
lines). This behavior contrasts with the findings of \citet{2006ApJ...647..483L}
and \citet{2013ApJ...764...21C}, who reported increasing net production with
initial mass. This discrepancy arises from the different explosion
prescriptions: while their models tune the explosion parameters to ensure the
ejection of the entire mantle above the Fe-core regardless of initialmass, we
adopt a standardized thermal bomb model with a fixed energy of $10^{51}$\,erg
across all masses. For very massive stars ($M_{\rm ini} \gtrsim 50\,M_\odot$),
the increasing gravitational binding energy of the envelope leads to substantial
fallback, preventing the ejection of the deep C/He-shell layers where
\iso{Fe}{60} is synthesized. In contrast, a posteriori mass cut determination in
the \citet{2013ApJ...764...21C}, which focuses on \iso{Ni}{56} yields after
ensuring mantle ejection, effectively bypasses the significant fallback. 
In our $70\,M_\odot$ and $80\,M_\odot$ models, the ejecta is dominated by the
outermost layers, whereas the \iso{Fe}{60}-rich regions almost entirely fall back
onto the remnant. This retention of \iso{Fe}{60} within the compact remnant in
the high-mass regime effectively prevents these stars from dominating the
integrated Galactic signal, even under top-heavy IMF assumptions.

It is worth noting, however, that our 1D hydrostatic treatment neglects
multi-dimensional mixing processes, which naturally occur during shock
propagation. In realistic multi-dimensional explosions, turbulent mixing can
transport material from the deep interiors into the outer envelope before
fallback occurs. Such mixing would allow a fraction of the synthesized
\iso{Fe}{60} to be dredged up and ejected into the interstellar medium, thereby
potentially increasing the integrated galactic yields and the resulting flux
ratio.

This leads to a bifurcated conclusion regarding the mechanism of yield
reduction. For lower to intermediate mass stars ($M_{\rm ini} \lesssim
50\,M_\odot$), the net loss of \iso{Fe}{60} during the explosion is primarily
governed by nuclear destruction via photodisintegration at the shock front.
Conversely, for the high-mass regime ($M_{\rm ini} > 50\,M_\odot$), the fallback
of deep, \iso{Fe}{60}-rich layers onto the compact remnant becomes the dominant
factor exacerbating the yield suppression. This dual-suppression mechanism
ensures that the Galactic \iso{Fe}{60}\iso{Al}{26} ratio remains robustly
anchored by the nucleosynthetic signature of the intermediate-mass population in
our models.

\subsection{Comparison with Galactic Observations}
\label{subsec:discussion_2}

Interpreting the agreement between our models and observations requires careful
consideration of the data analysis methods. While the galactic distribution of
\iso{Al}{26} is well-mapped, the specific morphology of \iso{Fe}{60} emission
remains poorly constrained \citep{2021PASA...38...62D}. Deriving a flux ratio
from current data requires assuming a spatial tracer for the \iso{Fe}{60}
signal. Previous studies, and our work, often assume that \iso{Al}{26} traces
\iso{Fe}{60} due to their shared origins in massive stars. However,
\citet{2020ApJ...889..169W} demonstrated that different tracer maps lead to
varying flux ratio estimates. Notably, they found that the \iso{Al}{26} map is
not the most statistically significant tracer for \iso{Fe}{60}, suggesting that
the two isotopes may have distinct spatial distributions.

Synthesizing results from various tracers, \citet{2020ApJ...889..169W}
constrained the galactic flux ratio to the range of $0.2\text{--}0.4$. In this
context, our DR results ($\sim 0.32$) fall comfortably within this window,
while the NR result ($\sim 0.18$) lies slightly below the lower bound. However,
if considering the exponential disk model which yields $0.184 \pm 0.042$, our NR
prediction offers an excellent match. This comparison highlights that the
tension between theory and observation is not solely a nuclear physics problem;
it also stems from astrophysical and observational systematics. 

Addressing these challenges requires a new generation of MeV $\gamma$-ray
observatories with significantly improved sensitivity and spatial resolution.
Upcoming and proposed missions offer a promising path to resolve these
discrepancies. Specifically, The Compton Spectrometer and Imager project
\citep[COSI,][]{2022icrc.confE.652T} is designed to provide high-resolution mapping of
diffuse isotopic emission, aiming to perform the first mapping of
\iso{Fe}{60}. Furthermore, proposed observatories such as
\citep[e-ASTROGAM,][]{2018JHEAp..19....1D} aim to provide detailed mapping of short-lived
radioactivities, including both \iso{Al}{26} and \iso{Fe}{60}, while the All-sky
Medium Energy $\gamma$-ray Observatory \citep[AMEGO,][]{2019BAAS...51g.245M} will
offer advanced line sensitivity for nuclear lines such as the $1.8$\,MeV emission
from \iso{Al}{26}. These missions represent essential improvements in
$\gamma$-ray instrumentation, providing the precision necessary to strictly
constrain the key nucleosynthetic observable of the \iso{Fe}{60}/\iso{Al}{26}
flux ratio.

\subsection{Systematic Uncertainties}
\label{subsec:discussion_3}

Finally, we address the limitations inherent in our numerical approach and the
physical assumptions adopted in this study.

\subsubsection{Numerical Sensitivities}
\label{subsubsec:numerical_sensiticities}
A primary challenge in modeling massive star nucleosynthesis is the sensitivity
of the stellar structure to numerical resolution and physical assumptions. As
noted in Section \ref{subsubsec:resolution}, we established a ``stable island''
in the parameter space where the pre-supernova structure converges robustly.
\rev{This approach effectively mitigates numerical instabilities during the advanced burning stages. The robustness of this numerical framework is validated by the consistency of our \iso{Al}{26} yields across different resolution sets, as well as the tight linear correlation maintained between the initial mass and the helium core mass. }
However, models around $15\,M_\odot$ remain numerically sensitive, 
particularly regarding the temperature profiles of the inner shells (see Figure
\ref{fig:resolution_2}), which directly dictates the efficiency of \iso{Fe}{60}
synthesis. While we have demonstrated that the observed suppression of
\iso{Fe}{60} is a robust physical signal rather than a numerical artifact, the
fluctuations in our control yields (e.g., \iso{Al}{26}) suggest a persistent
numerical noise floor of $\sim 0.26$\,dex in absolute yield values.

\subsubsection{Stellar Physics and Explosion Prescription}
\label{subsubsec:stellar_physics}
A primary source of systematic uncertainty stems from the 1D treatment of
internal mixing. Our convection prescription relies on the 1D Mixing Length
Theory, whereas convection in massive stars is inherently multidimensional.
Consequently, while the relative suppression of \iso{Fe}{60} due to the updated
decay rate is a robust physical effect, the absolute yield values carry
uncertainties inherent to 1D stellar evolution. Furthermore, stellar rotation,
omitted in our current grid, is generally found to increase \iso{Al}{26} yields
due to enhanced mixing and mass loss, while also impacting \iso{Fe}{60}
production \citep{2012A&A...537A.146E, 2018ApJS..237...13L, 2021ApJ...923...47B,
2025ApJ...991...21F}.

Our explosion modeling is also simplified, adopting a uniform thermal bomb
approach with a fixed energy of 1\,foe. In reality, the explosion mechanism is
still not fully understood, and the explosion energy likely scales with
progenitor properties, which would in turn modify the shock-induced
nucleosynthesis and the extent of fallback. This simplification likely
contributes to the discrepancies in net explosive yields observed in our grid
compared to previous studies. For instance, we do not observe the consistent net
explosive production of \iso{Fe}{60} and \iso{Al}{26} across the entire mass
range reported by \citet{2006ApJ...647..483L}. 
\rev{Specifically, this leads to the mass-dependent bifurcation identified in our yields: for stars below $\sim 50\,M_\odot$, the reduction in \iso{Fe}{60} is primarily a result of shock-induced photodisintegration at the shell base. However, for more massive stars ($M > 50\,M_\odot$), the suppression is significantly dominated by fallback, as the increasing gravitational binding energy of the progenitor exceeds our assumed 1\,foe explosion energy.}
While a comprehensive exploration
of explosion physics is beyond the scope of this work, we emphasize that the
systematic suppression of \iso{Fe}{60} in our models originates primarily in the
hydrostatic C-shell. Therefore, this effect is expected to persist as a dominant
correction regardless of the specific explosion model or energy adopted.

\subsubsection{Incompleteness of Stellar Sources}
\label{subsubsec: stellar sources}

Our current grid covers $14\text{--}80\,M_\odot$, leaving other potential
sources and physical processes unaccounted for. First, we do not model the
lower-mass end of massive stars ($8\text{--}13\,M_\odot$). These stars are more
numerous and are expected to produce a higher yield of \iso{Al}{26} relative to
\iso{Fe}{60}\citep{2006ApJ...647..483L, 2021ApJ...923...47B}, potentially
lowering the integrated flux ratio further. Second, Asymptotic Giant Branch
(AGB) stars are known contributors, though their \iso{Fe}{60} yields remain
highly uncertain, and both \iso{Al}{26} and \iso{Fe}{60} yields from AGB stars
are considered to be subdominant compared to massive stars
\citep{2008NewAR..52..416L, 2010MNRAS.403.1413K, 2018PrPNP.102....1L}. 

On the high-mass end ($70\text{--}80\,M_\odot$), our models predict significant
fallback, resulting in direct collapse to black holes and negligible
contributions to the final yields. 
However, even for stars that fail to explode,
a significant amount of \iso{Al}{26} can be ejected via strong stellar winds
during the Wolf-Rayet phase \citep{2005A&A...429..613P}. In contrast,
\iso{Fe}{60} is synthesized in deeper layers and remains retained in the compact
remnant during fallback or collapse. This spatial decoupling would further
suppress the integrated Galactic \iso{Fe}{60}/\iso{Al}{26} ratio. 
We also note that we did not perform a self-consistent "explodability" analysis
(e.g., based on compactness parameters)\citep{2011ApJ...730...70O,
2014ApJ...783...10S, 2016ApJ...818..124E}; the precise mass limits for
successful supernovae remain an open question which could systematically shift
the predicted flux ratio.

\subsubsection{Nuclear Data Uncertainties}
\label{subsubsection: nuclear uncertainties}
Beyond the \iso{Fe}{59} decay rate, uncertainties in other key reaction rates
persist. Recently, \citet{2024NatCo..15.9608S} provided the most complete
estimate on the \iso{Fe}{59}$(n, \gamma)$\iso{Fe}{60} reaction using the
$\beta$-Oslo method, which is 1.6 \text{--} 2.1 times higher than the previous
estimation. While this significantly reduces the previously large uncertainty
($\sim 40\%$) associated with the production channel, it also implies an even
higher potential for \iso{Fe}{60} overproduction. Consequently, the final
\iso{Fe}{60} abundance remains critically sensitive to the balance between this
neutron capture and the \iso{Fe}{59} $\beta^-$-decay rate during shell burning.
Additionally, for the decay rate of \iso{Fe}{60} in stellar environments,
discrepancies still exist between LMP and much earlier estimates
\citep{1982ApJ...252..715F}. 

Meanwhile, the helium burning rate, the triple-$\alpha$ and \iso{C}{12}$(\alpha,
\gamma)$\iso{O}{16}, introduces significant variations.
\citet{2010ApJ...718..357T} demonstrated that these reaction rates influence the
advanced burning stages and final yields in a complex manner; specifically,
varying these rates within their $1\sigma$ uncertainty limits can lead to a
factor of $3\text{--}5$ variation in the final \iso{Fe}{60} yield, whereas the
\iso{Al}{26} yield is notably less sensitive. While adjusting this rate could
also align models with observations, it affects the entire structural evolution.
\rev{\citet{2013ApJ...762...31P} also reported that the uncertainty in the \iso{C}{12}+\iso{C}{12} reaction rate significantly influences the pre-explosive synthesis of \iso{Fe}{60} and \iso{Al}{26}.}
Furthermore, neutron source reactions, such as \iso{Ne}{22}$(\alpha,
n)$\iso{Mg}{25}, are also crucial
for overall \iso{Fe}{60} yields and carry uncertainties. In contrast, the update
to the \iso{Fe}{59} $\beta^-$-decay rate provides a localized correction that
systematically suppresses \iso{Fe}{60} without requiring extreme variations in
stellar structure.

Despite these remaining uncertainties, the crucial finding of this work stands:
the updated \iso{Fe}{59} $\beta^-$-decay rate introduces a systematic and
significant suppression of \iso{Fe}{60} production that is robust across the
relevant mass range. \rev{By incorporating this updated nuclear physics, we
identify a pivotal mechanism that significantly mitigates the long-standing
overestimation of the galactic \iso{Fe}{60}/\iso{Al}{26} flux ratio. This
contribution moves theoretical predictions closer to the current observational
limits, although a definitive resolution of the discrepancy will likely require
the combined consideration of all newly available nuclear data across multiple
reaction channels.}

\section{Conclusions}
\label{sec:conclusion}

In this work, we performed a systematic investigation into the nucleosynthesis
of the radioisotopes \iso{Al}{26} and \iso{Fe}{60} in massive stars,
specifically focusing on the impact of the recently updated \iso{Fe}{59} stellar
$\beta^-$-decay rate. Using the \texttt{MESA} code, we constructed a grid of
stellar evolution models ranging from $14$ to $80\,M_\odot$, coupled with a
rigorous sensitivity analysis to ensure numerical stability. Our main
conclusions are summarized as follows:

\begin{itemize}
     \item[-] Significant Suppression of \iso{Fe}{60}:
     
     Implementing the experimentally constrained \iso{Fe}{59} decay rate
     \citep{2021PhRvL.126o2701G} fundamentally alters the s-process path in
     massive stars. By enhancing the depletion of \iso{Fe}{59} via
     $\beta^-$-decay, the new rate competes effectively with the neutron capture
     channel, particularly during convective Carbon shell burning. This leads to
     a global reduction in \iso{Fe}{60} yields by approximately $0.28$\,dex
     ($\sim 47\%$) compared to models using the default LMP rate. This result
     suggests that previous nucleosynthesis models relying on theoretical rates
     have systematically overestimated the galactic \iso{Fe}{60} production.

     \item[-] \rev{Consistency} with Galactic Observations:
     
     The suppressed \iso{Fe}{60} yields from our NR models lead to a predicted
     Galactic $\gamma$-ray flux ratio of $I(\iso{Fe}{60})/I(\iso{Al}{26})
     \approx 0.18$. This value aligns closely with the observational constraint
     of $0.184 \pm 0.042$ derived from INTEGRAL data
     \citep{2020ApJ...889..169W}. 
     \rev{ However, as discussed in Section
     \ref{subsubsection: nuclear uncertainties}, the inclusion of the newly
     available experimental update for the $^{59}\text{Fe}(n, \gamma)$ capture
     rate \citep{2024NatCo..15.9608S}, alongside future experimental constraints
     on other uncertain reactions, will provide further refinement to this
     theoretical prediction. Nevertheless, the observed weak dependence of this
     ratio on the IMF slope provides a robust theoretical baseline, suggesting
     that the $\beta^-$ decay enhancement remains a key factor in mitigating the
     long-standing tension between theoretical yields and observational limits.}

     \item[-] Bifurcated Reduction Mechanism:
     
     Contrary to previous studies where \iso{Fe}{60} yields were often found to
     increase monotonically during the explosion, we identified a mass-dependent
     bifurcation in the yield reduction mechanism, 
     \rev{under the assumption of
     a uniform explosion energy of $1.0 \times 10^{51}$ erg (1\,foe) across our
     model grid.} 
     For intermediate-mass stars ($M \lesssim 50\,M_\odot$), the
     yield reduction is primarily driven by shock-induced photodisintegration at
     the shell base. Conversely, for high-mass stars ($M > 50\,M_\odot$), the
     suppression is dominated by significant fallback, 
     \rev{ as the increased
     gravitational binding energy outpaces the fixed explosion energy. While the
     identified bifurcation is framed within our current explosion modeling (see
     Section \ref{subsubsec:stellar_physics}), the potential impacts of varied
     explosion energies and advanced explodability criteria on the final
     \iso{Fe}{60} and \iso{Al}{26} yields remain compelling subjects for future
     systematic investigations. }

\end{itemize}


\begin{acknowledgments}
This work was supported by the National Natural Science
Foundation of China under grant No.12588202, National Key R\&D Program of
China under grant No.2024YFA1611900, and Strategic Priority Research Program of
Chinese Academy of Sciences under grant No.1160102.
We thank Bingshui Gao for generously providing the experimental data on the
\iso{Fe}{59} $\beta^-$ decay rate used in this work. 

\end{acknowledgments}

\section*{Software and Data Availability} 
\rev{ The MESA configuration files (inlists), customized nuclear network, and
experimental reaction rate tables required to reproduce the stellar evolution
and nucleosynthesis simulations presented in this work are publicly available on
Zenodo via the AAS Journals Community at
\dataset[doi:10.5281/zenodo.20177425]{https://doi.org/10.5281/zenodo.20177425}.
This Zenodo deposit also includes the stellar model properties (Table
\ref{tab:basic_evolution}), isotopic yields (Table \ref{tab:yields}), and the
Python-based analysis routines used for convolving yields with the IMF and
generating Figure \ref{fig:flux_ratio}. For interactive viewing convenience, the
analysis scripts and tabulated data are also mirrored on GitHub at
\url{https://github.com/Tan-0321/Fe59-decay-impact-Fe60-Al26}. }


\appendix

\section{Resolution Sensitivity and Numerical Robustness}
\label{sec:appendix_resolution}

To ensure the robustness of our nucleosynthetic yields, we systematically
investigated the impact of numerical resolution on core properties and stellar
structure. This appendix details our convergence strategy and validates that our
production models (using the extensive \texttt{mesa\_162.net} network) reside
within the numerically stable regime established by a dedicated test grid.

\subsection{Methodology and Convergence Criteria}
\label{app:methodology}

We constructed a test grid covering initial masses of $15\,M_{\odot}$,
$25\,M_{\odot}$, and $40\,M_{\odot}$ using the simplified
\texttt{approx21\_cr60\_plus\_co56.net} network. For each mass, we varied two
key resolution control parameters: the maximum mass fraction change per cell
(\texttt{max\_dq}) and the temporal resolution control
(\texttt{delta\_lgRho\_center\_limit}).

\subsubsection{Phase I: Helium Core Convergence}
\begin{figure*}[htbp]
\centering
\includegraphics[width=1.0\textwidth]{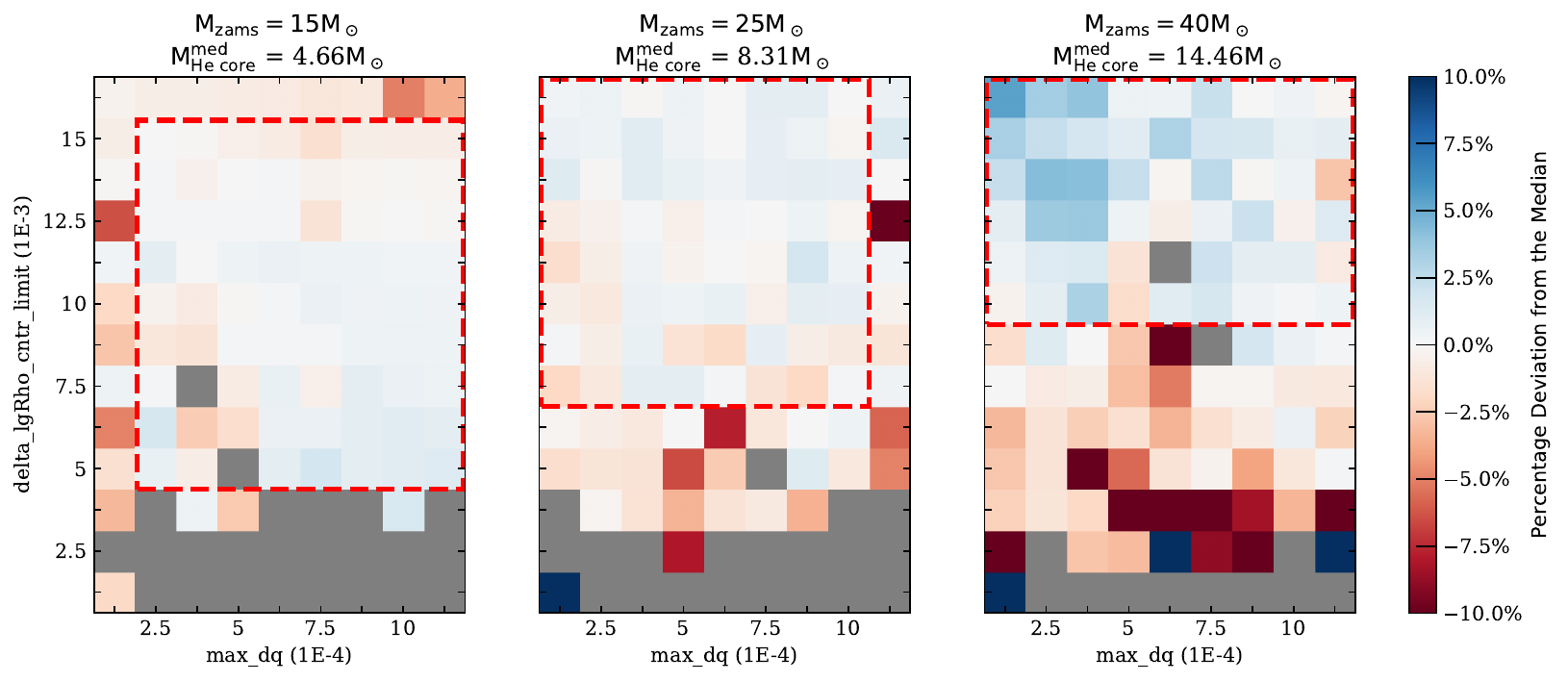}
\caption{Percentage deviation of the helium core mass from the median value for models with $15, 25,$ and $40\,M_\odot$. The red dashed rectangle highlights the region where deviation is $<5\%$. Grey blocks indicate models that failed due to excessively small timesteps or convergence issues.}
\label{fig:resolution_1}
\end{figure*}

For the evolution from ZAMS to helium depletion, we focus on the convergence of
the helium core mass ($M_{\text{He}}$). We define the percentage deviation as:

\begin{equation}
\delta M_{\text{He}} = \frac{M_{\text{He}}(\Delta m, \Delta t) - M_{\text{He}}^{\text{ref}}}{M_{\text{He}}^{\text{ref}}} \times 100\%
\label{eq:dev}
\end{equation}

where $M_{\text{He}}^{\text{ref}}$ is the median value across the resolution
grid. As shown in Figure~\ref{fig:resolution_1}, the timestep control
(\texttt{delta\_lgRho\_center\_limit}) is the dominant factor. We adopted a
threshold of $5\%$ (red dashed rectangle) to define our baseline resolution for
the production run.

\subsubsection{Phase II: Pre-Supernova Thermal Structure}
\begin{figure*}[htbp]
     \centering
     \includegraphics[width=1.0\textwidth]{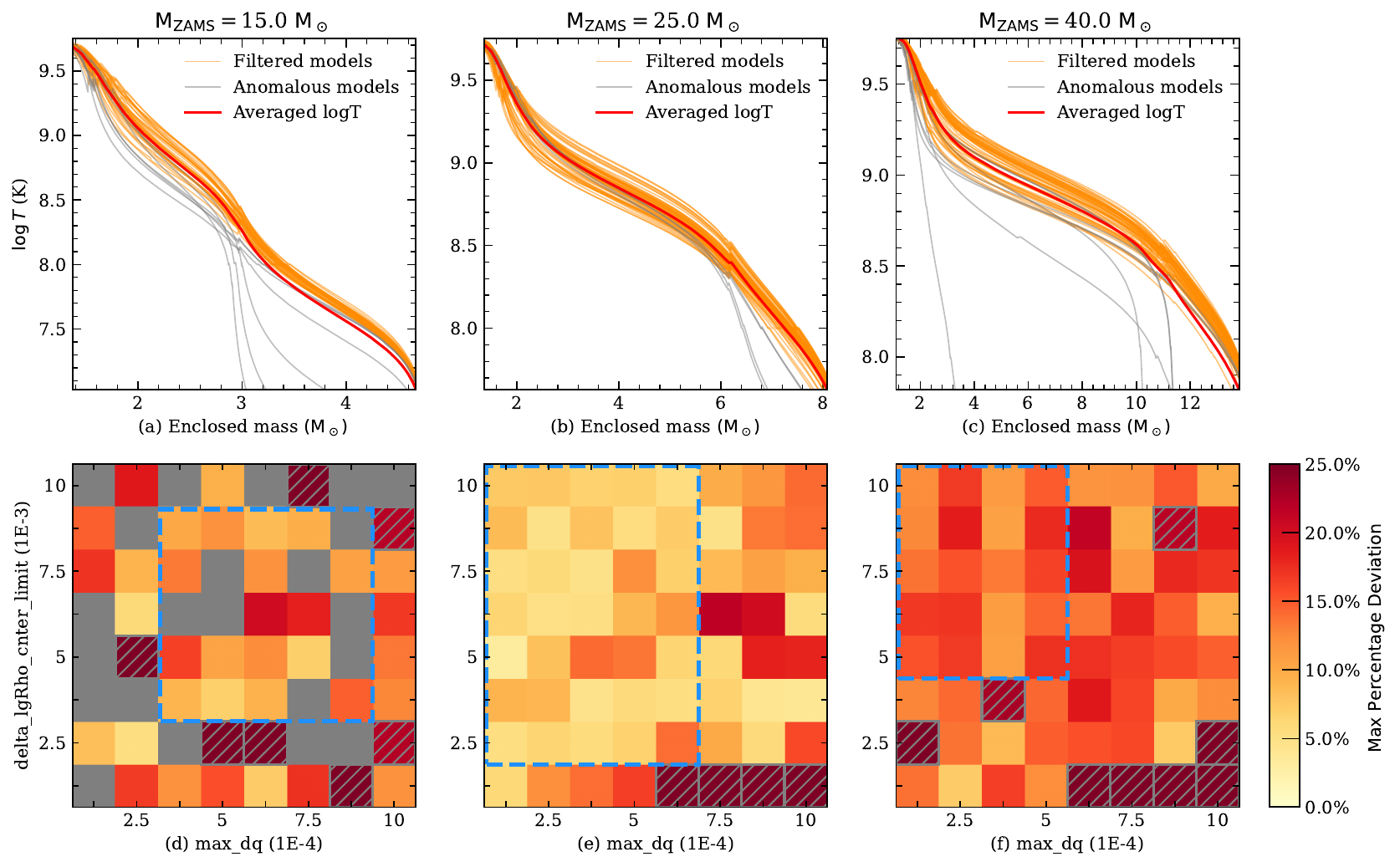}
     \caption{Resolution sensitivity analysis for advanced burning stages. \textbf{Top:} Temperature profiles prior to iron core infall. Grey curves are outliers ($>50\%$ deviation), orange are valid models, and red is the average. \textbf{Bottom:} Map of maximum temperature deviation ($D^{\text{max}}_{\text{temp}}$). The blue dashed rectangles indicate the chosen stability parameters for production models.}
     \label{fig:resolution_2}
\end{figure*}

For the advanced burning stages (post-helium depletion), we prioritize the
stability of the temperature profile. We define the maximum absolute percentage
deviation of the temperature profile just before iron core infall ($\log T_c
\approx 9.85$) as:

\begin{equation}
D^{\text{max}}_{\text{temp}} = \max_{m} \left\{ 
     \left|
     \frac{T(m) - \bar{T}(m)}{\bar{T}(m)} 
     \right|
\times 100\%
     \right\}
\label{eq:temp_dev}
\end{equation}

where $\bar{T}(m)$ is the refined average profile. Figure~\ref{fig:resolution_2}
(bottom panels) shows the stability map. While $25$ and $40\,M_{\odot}$ models
show broad stability windows, the $15\,M_{\odot}$ models are more stochastic. We
selected resolution settings within the blue dashed ``stability islands"
($D^{\text{max}}_{\text{temp}} < 20\%$) for our production runs.

\subsection{Validation of Production Models}
\label{app:validation}

A critical step is to confirm that the resolution criteria derived from the test
grid (small network) remain valid for the production grid (large network).

\paragraph{Structural Consistency}
In Figure~\ref{fig:hecore_compare} (main text), we demonstrated that the helium
core masses of our production models align perfectly with the test grid
predictions and literature values.

\begin{figure*}[htbp]
     \centering
     \includegraphics[width=1.0\textwidth]{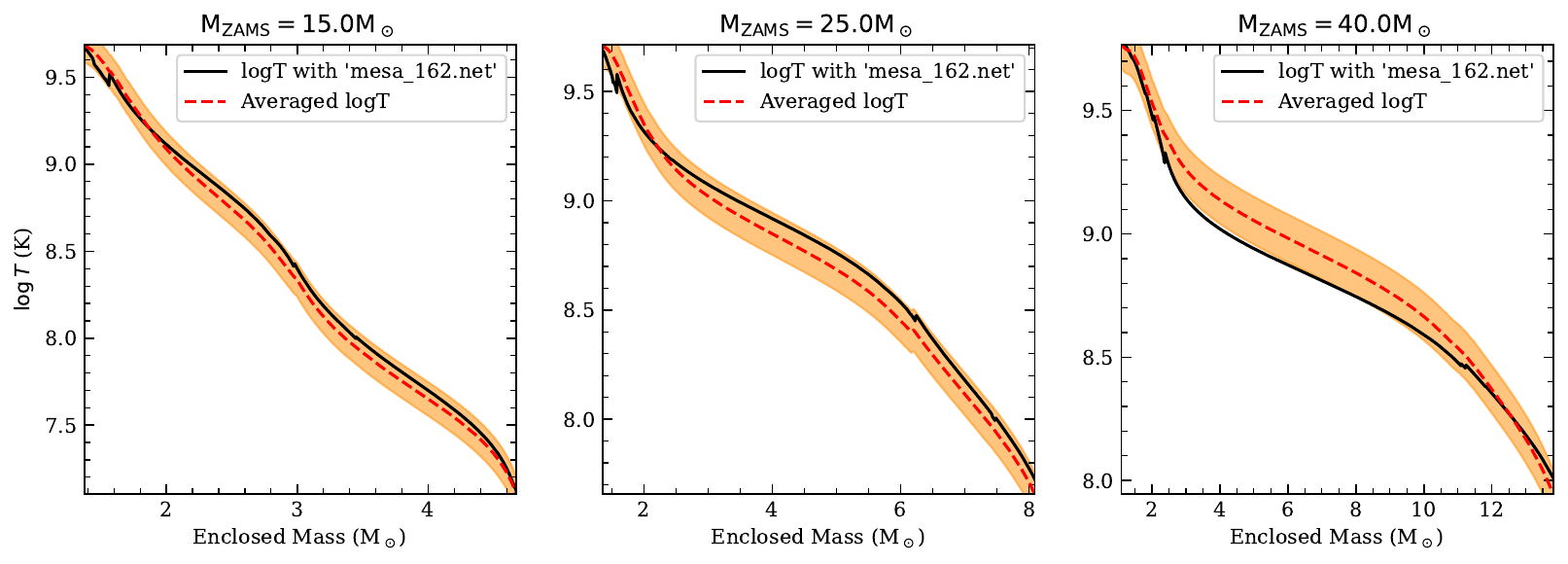}
     \caption{Validation of production models. Black lines show the temperature 
     profiles of our production models ($15, 25, 40\,M_\odot$) using the large 
     network. Red dashed lines are the averaged profiles from the resolution 
     test grid. Orange shaded regions indicate the $\pm 20\%$ deviation 
     tolerance. Note the excellent agreement, confirming that network size does 
     not induce significant structural drift.}
     \label{fig:temp_compare}
\end{figure*}

\paragraph{Thermal Profile Consistency}
Figure \ref{fig:temp_compare} compares the pre-supernova temperature profiles of
the production models (using \texttt{mesa\_162.net}) against the averaged
profiles from the test grid. For the 15 and 25 $M_{\odot}$ models, the
production profiles fall well within the $20\%$ deviation envelope (orange
shaded region). The 40 $M_{\odot}$ model shows a minor excursion near the C/Ne
burning shell interface ($\sim 4\,M_{\odot}$), which is a known region of high
sensitivity to network-convection coupling. However, the overall thermal
structure is preserved. This confirms that our two-stage resolution strategy
provides a robust physical foundation for the nucleosynthesis analysis presented
in this work.

\bibliography{sample7}{}
\bibliographystyle{aasjournal}



\end{document}